%% file: main.tex
\documentclass[sigconf,nonacm ]{acmart}

\usepackage{amsmath,amsfonts,amsthm}
\usepackage{booktabs}
\usepackage{multirow}
\usepackage{subfigure}
\usepackage{hyperref}
\usepackage{url}
\usepackage{graphicx}
\usepackage{textcomp}
\usepackage{xcolor}
\usepackage{float}
\usepackage{algorithm}
\usepackage{algorithmic}
\usepackage{caption}
\begin{document}
\sloppy
\title{Model Extraction and Defenses \\ on Generative Adversarial Networks}
\author{Hailong Hu}
\affiliation{\institution {University of Luxembourg}
  \city{~}
  \country{~}
}
\author{Jun Pang}
\affiliation{\institution {University of Luxembourg}
  \city{~}
  \country{~}
}

\begin{abstract}
Model extraction attacks aim to duplicate a machine learning model through query access to a target model. Early studies mainly focus on discriminative models. Despite the success, model extraction attacks against generative models are less well explored. In this paper, we systematically study the feasibility of model extraction attacks against generative adversarial networks~(GANs). Specifically, we first define accuracy and fidelity on model extraction attacks against GANs. Then we study model extraction attacks against GANs from the perspective of accuracy extraction and fidelity extraction, according to the adversary's goals and background knowledge. We further conduct a case study where an adversary can transfer knowledge of the extracted model which steals a state-of-the-art GAN trained with more than 3 million images to new domains to broaden the scope of applications of model extraction attacks. Finally, we propose effective defense techniques to safeguard GANs, considering a trade-off between the utility and security of GAN models.
\end{abstract}



\maketitle
\thispagestyle{plain}
\pagestyle{plain}

\input{1_introduction}
\input{2_related_work}
\input{3_preliminaries}
\input{4_Taxonomy}

\input{5_accuracy}

\input{6_fidelity}

\input{7_case_study}
\input{8_defense}

\input{9_conclusion}

\begin{acks} 
This work is supported by the National Research Fund, Luxembourg~(Grant No. 13550291).
\end{acks}

%
\clearpage
\bibliographystyle{ACM-Reference-Format}
\bibliography{reference}

\appendix
\input{appendix}
\end{document}

%% file: 1_introduction.tex
\section{Introduction}
\label{sec:intro}
Over the past few years, machine learning, deep learning in particular, has gained significant advances in a variety of areas, such as computer vision and natural language processing (NLP). For instance, recently FixEfficientNet-L2~\cite{touvron2019fixing} has achieved 98.7\% accuracy on the ImageNet recognition benchmark~\cite{russakovsky2015imagenet}, compared to the accuracy of 73.8\% in 2011. NLP models created by fine-tuning a pretrained model, i.e. BERT~\cite{devlin2018bert}, also become extremely successful in many natural language tasks~\cite{nogueira2019passage,liu2019fine,lee2020biobert}.
Generative adversarial networks (GANs) are able to generate photo-realistic images that humans are difficult to distinguish~\cite{BigGAN2018large,StyleGAN12019style,StyleGAN22019analyzing}. In general, these state-of-the-art models are often considered as the intellectual property of model owners and are closely safeguarded. The reasons are from at least two aspects. First, obtaining a practical deep learning model is non-trivial. This is because training a model requires a large number of training data, intensive computing resources and human resources~\cite{russakovsky2015imagenet,yu2015lsun,touvron2019fixing,devlin2018bert,GPT3brown2020language,StyleGAN22019analyzing}. Second, deep learning models themselves are confidential, and exposure to deep learning models to potential adversaries poses a threat to security and privacy~\cite{lowd2005adversarial,papernot2017practical,shokri2017membership,salemml,tramer2016stealing,luvcic2019high}. However, model extraction attack --- a novel attack surface targeting at duplicating a model only through query access to a target model, has recently emerged and gained significant attention from the research community.

In the early study, Tram{\`e}r et al.~\cite{tramer2016stealing} first attempt model extraction on traditional machine learning models and shallow neural networks, such as logistic regression, decision tree, support vector machine and multilayer perceptrons. Since then, Jagielski et al.~\cite{jagielski2019high} further mount the attack against a million of parameters model trained on billions of Instagram images~\cite{Mahajan_2018_ECCV}, which makes model extraction attack more practical. In addition to model extraction on deep convolutional neural networks about image classification, there are some works studying the problem of model extraction in NLP tasks~\cite{Krishna2020Thieves,takemura2020model}. For instance, with the assumption that victim models are trained based on the pretrained BERT model, Krishna et al.~\cite{Krishna2020Thieves} show that an adversary can effectively extract language models whose performance is only slightly worse than that of the victim models. However, to the best of our knowledge, these model extraction attacks mainly focus on discriminative models. The attack against generative models, GANs in particular, is still an open question.

Comparing to model extraction attacks on discriminative models, we observe that there exist some differences for generative models. First, adversaries can leverage output information from target models such as labels, probabilities and logits, to mount model extraction attacks on discriminative models~\cite{lowd2005adversarial,orekondy2019knockoff,jagielski2019high,tramer2016stealing}, while generative models do not provide such information but only return images. 
Second, for model extraction attacks on discriminative models, they are evaluated by a test dataset. In contrast, unsupervised generative models aiming to learn the distribution of training data are evaluated by quantitative measures such as Fr{\'e}chet Inception Distance (FID)~\cite{FIDl2017gans} and multi-scale structural similarity (MS-SSIM)~\cite{odena2017conditional}, or qualitative measures such as preference judgment~\cite{huang2017stacked,zhang2017stackgan}.
Therefore, these differences indicate that model extraction strategies, evaluations and defenses on generative models are very different from these on discriminative models. 

In this paper, we aim to systematically study the feasibility of model extraction attacks against GANs from the perspective of accuracy extraction and fidelity extraction. First, we define {\it accuracy} and {\it fidelity} of model extraction on GANs. More specifically, when an adversary mounts model extraction attacks against GANs, {\it accuracy} measures the difference of data distribution between the attack model and the target model, while {\it fidelity} ensures the distribution of the attack model is consistent with the distribution of the training set of the target model.
In the next step, according to the adversary's goals and the background information that they can have access to (see Figure~\ref{fig:Attack settings}), we systematically study two different types of attacks on GANs: accuracy extraction attack and fidelity extraction attack, which are shown in Figure~\ref{fig:attack types}.

\textbf{Accuracy extraction attack.} Adversaries mounting accuracy extraction focus on {\it accuracy} and they aim to steal the distribution of a target model. 
For this attack, we assume adversaries have no knowledge of the architecture of target models, and they either obtain a batch of generated data that the model owner has publicly released or query the target model to obtain generated data. It can be considered as a black-box accuracy attack. After obtaining the generated data, adversaries can train a copy of the target GAN model. We study two different target models: Progressive GAN~(PGGAN)~\cite{PGGAN2018progressive} and Spectral Normalization GAN~(SNGAN)~\cite{SNGAN2018spectral}. Extensive experimental evaluations show that accuracy extraction can achieve an acceptable performance with only about 50k queries~(i.e., 50k generated samples).
When we continue to increase the number of queries, we find that it cannot bring significant improvement of the {\it fidelity} of attack models. 
This is mainly because the discriminator of a target GAN model is often better than its corresponding generator and it is very hard to reach global optimum~\cite{DRS2018discriminator}. In other words, directly querying the target model enables the attack model to be more consistent with the target generator rather than the real data distribution of the target model~(see Figure~\ref{fig:toy_example} for an example). Therefore, it motivates us to perform fidelity extraction to improve the {\it fidelity} of attack models.

\textbf{Fidelity extraction attack.} Adversaries mounting fidelity extraction concentrate on {\it fidelity} and they target at stealing the distribution of the training set of a target model.
In order to mount a fidelity extraction attack, we assume that adversaries can obtain more background knowledge. We discuss two subcategories for fidelity extraction: partial black-box fidelity extraction and white-box fidelity extraction. For the former, we assume that adversaries are able to have access to partial real data from the target GAN models, in addition to querying the target model. Extensive experimental evaluations show that adding partial real data is an effective method in terms of improvement of {\it fidelity} of attack models. For the latter, we assume adversaries can obtain the discriminator from the target GAN model and partial real data, which is the most knowledgeable setting. Also, adversaries can query the target model. We utilize the discriminator to subsample generated samples. These refined samples are more close to real data distribution, compared to samples are directly generated by the target model~(see Figure~\ref{fig:MH_samples}). Then, we use these refined samples and partial real data to train our attack model. Extensive experimental evaluations also show improvement of the {\it fidelity} of attack models. 

\textbf{Case study.} We perform one case study to further demonstrate the impact of model extraction attacks on a large-scale scenario. In this case study --- model extraction based transfer learning, we show that stealing a state-of-the-art GAN model can enable adversaries to enhance the performance of their own GAN model by transfer learning. Compared with training from scratch on LSUN-Classroom dataset with 20.34 FID~\cite{PGGAN2018progressive}, model extraction based transfer learning achieves 16.47 FID, which is the state-of-the-art performance on LSUN-Classroom dataset. 

\textbf{Defense.} Both accuracy extraction and fidelity extraction attacks on GANs compromise the intellectual property of model providers. In particular, fidelity extraction aiming to steal the distribution of the training set of a target model can further severely breach the privacy of the training set.
Therefore, we propose possible defense techniques by considering two aspects: $\it accuracy$ and $\it fidelity$. In terms of $\it accuracy$ of model extraction, limiting the number of queries is an effective method. In terms of $\it fidelity$ of model extraction,
we believe that a high fidelity attack model requires adversaries to have access to generated data which can be representative for real data distribution. The performance of model extraction attacks will be attenuated if adversaries only obtain a partial or distorted distribution of generated data. Thus, we propose two types of perturbation-based defense strategies: input and output perturbation-based approaches, to reveal less distribution information by increasing the similarity of samples or lowering the quality of samples~\cite{agustsson2018optimal}. The input perturbation-based approaches include linear and semantic interpolation perturbation while the output perturbation-based approaches include random noise, adversarial example noise, filtering and compression perturbation. 
Extensive experimental evaluations show that, compared to queries from the prior distribution of the target model, the equal amount of queries by perturbation-based defenses can effectively degrade the $\it fidelity$ of attack models.

\smallskip\noindent
\textbf{Summary of contributions.} Our contributions in the current work are threefold:
\begin{enumerate}
	\item we conduct the first systematic study of model extraction attacks against GANs and devise accuracy extraction attacks and fidelity extraction attacks for GANs;
	\item we preform one case study to illustrate the impact of model extraction attacks against GANs on a large-scale scenario;
	\item we propose new effective defense measures to mitigate model extraction attacks against GANs. 
\end{enumerate}

\smallskip\noindent
\textbf{Organization.} The rest of the paper is organized as following. The next section~\ref{sec:related} reviews related work. Section~\ref{sec:pre} introduces the preliminary knowledge, and Section~\ref{sec:taxonomy} taxonomizes the space of model extraction attacks on GANs. Section~\ref{sec:accuracy} and Section~\ref{sec:fidelity} introduce the accuracy extraction and fidelity extraction, respectively. Section~\ref{sec:casestudy1} presents one case study. In Section~\ref{sec:defenses}, we discuss possible defense mechanisms. Section~\ref{sec:conclusion} concludes this paper.

%% file: 2_related_work.tex
\section{Related work}
\label{sec:related}
\noindent
\textbf{{Generative adversarial networks (GANs).}} GANs have achieved impressive performance in a variety of areas, such as image synthesis~\cite{DCGAN2015unsupervised,WGAN2016improved,SNGAN2018spectral,PGGAN2018progressive,BigGAN2018large,StyleGAN12019style,StyleGAN22019analyzing,lin2019coco}, image-to-image translation~\cite{zhu2016generative,park2019semantic,liu2019few}, and texture generation~\cite{li2016precomputed,xian2018texturegan}, since a framework of GAN was first proposed by Goodfellow et al. in 2014~\cite{goodfellow2014generative}. For image synthesis tasks, the current state-of-the-art GANs~\cite{SNGAN2018spectral,PGGAN2018progressive,BigGAN2018large,StyleGAN12019style} are able to generate highly realistic and diverse images. For instance, SNGAN~\cite{SNGAN2018spectral} generates realistic images by a spectral normalization method to stabilize the training process. PGGAN~\cite{PGGAN2018progressive} proposed by Karras et al. is the first GAN that successfully generates real-like face images at a high resolution of 1024 $\times$ 1024, applying a progressive training strategy. Unlike the PGGAN training in an unsupervised method, BigGAN~\cite{BigGAN2018large} proposed by Brock et al. aims to generate high-quality images from a multi-class dataset by conditional GANs which leverage information about class labels. Recently, StyleGAN~\cite{StyleGAN12019style} has further improved the performance of GANs on high-resolution images through adding neural style transfer~\cite{huang2017arbitrary}. In this paper, \textit{we choose SNGAN and PGGAN as the target models to be attacked by model extraction, considering their 
impressive performance on image generation. StyleGAN is also used as a target model in a case study in Section~\ref{sec:casestudy1}}.

\smallskip\noindent
\textbf{{Model extraction attacks.}} With the availability of machine learning as a service (MLaaS), model extraction attack has received much attention from the research community~\cite{tramer2016stealing,Krishna2020Thieves,carlini2020cryptanalytic,jagielski2019high,chen2020stealing}, which aims to duplicate~(i.e., `steal') a machine learning model. This type of attack can be categorized into two classes: accuracy model extraction and fidelity model extraction. In terms of accuracy model extraction, it was first proposed by Tram{\`e}r et al.~\cite{tramer2016stealing}, where the objective of the attack is to gain similar or even better performance on the test dataset for the extracted model. Since then, various methods attempting to reduce the number of queries have been developed for further improving the attack efficiency, such as model extraction using active learning~\cite{pal2020activethief,chandrasekaran2020exploring} or semi-supervised learning~\cite{jagielski2019high}. In terms of fidelity model extraction, it requires the attack model to faithfully reproduce predictions of the target model, including the errors which occur in the target model. Typical works include model reconstruction from model explanation~\cite{milli2019model}, functionally equivalent extraction~\cite{jagielski2019high} and cryptanalytic extraction~\cite{carlini2020cryptanalytic}.
In addition to model extraction attacks on images, there are several work about model extraction in natural language processing~\cite{Krishna2020Thieves,takemura2020model}. Krishna et al.~\cite{takemura2020model} mount model extraction attacks against BERT-based models and the performance of the extracted model is slightly worse than that of the target model. Overall, these studies mainly focus on discriminative models, such as regression and convolutional neural networks for classification, and recurrent neural networks for natural language processing. \textit{Unlike the existing studies, our work aims to study model extraction attacks against GANs. }

In addition to model extract attacks, there are other types of attacks in relation to privacy and security~\cite{he2019towards,Xudong2020Privacy,carlini2020extracting}, such as membership inference attacks~\cite{shokri2017membership,salemml,shokri2019privacy,hayes2019logan,chen2019gan} and property inference attacks~\cite{ganju2018property}. Some efforts have been also made to investigate membership inference attacks against GANs, where queries to a GAN model can reveal information about the training dataset~\cite{hayes2019logan,chen2019gan,hilprecht2019monte}. Overall, these studies mainly focus on privacy on the training dataset, \textit{while model extraction attacks in our paper concentrate on machine learning model itself.}

\smallskip\noindent
\textbf{{Model extraction defenses.}} Defense for model extraction can be broadly classified into two categories: restricting the information returned by models~\cite{tramer2016stealing,lee2018defending} and differentiating malicious adversaries from normal users~\cite{juuti2019prada}. Tram{\`e}r et al. propose a defense where the model should only return class labels instead of class probabilities~\cite{tramer2016stealing}. Recently, a technique PRADA has proposed to guard machine learning models by detecting abnormal query patterns~\cite{juuti2019prada}. Watermarking ML models as a passive defense mechanism recently has been proposed to claim model's ownership~\cite{jia2020entangled}. However, these defense techniques are used to protect discriminative models where models return probabilities or labels. In this paper, \textit{we focus on defense approaches safeguarding generative adversarial networks where models return images.} 

%% file: 3_preliminaries.tex
\section{Preliminaries}
\label{sec:pre}
In this section, we begin with the general structure of GANs. Then, we proceed with discussing model extraction attacks in a general machine learning setting. Finally, we describe datasets used in this paper.

\subsection{Generative adversarial networks}
GAN is a generative model where it adversarially learns the unknown true distribution $p_{r}$ on the training data $\mathbf X$. As shown in Figure~\ref{fig:Attack settings}, a GAN generally consists of two components: a generator $G$ and a discriminator $D$. $G$ is responsible for generating fake data $x_{g} = G\left(z\right)$, where the latent code $z$ is sampled from a prior distribution $p_{z}$, such as Gaussian distribution or uniform distribution, while $D$ takes the role of a binary classifier which differentiates real-like samples $x_{g}$ from real samples $x_{r} \in \mathbf{X}$ as accurately as possible. The seminal GAN~\cite{goodfellow2014generative} is trained through optimizing the following loss functions:
\begin{equation}
    L_D = -\mathbb{E}_{x\sim p_{r}}\left [ \log{D\left(x\right)} \right ] - \mathbb{E}_{z \sim p_{z}}\left [1 -  \log{D\left(G\left(z\right)\right)} \right ] 
\end{equation} 
\begin{equation}
    L_G = -\mathbb{E}_{z \sim p_{z}}\left [ \log{D\left(G\left(z\right)\right)} \right ]
\end{equation} 

If $D$ and $G$ converge and reach global equilibrium, then $p_{r}\left(x\right) = p_{g}\left(x\right)$, where $p_{g}\left(x\right)$ is the generator's distribution. For a fixed $G$, the optimal discriminator $D^{*}$ can be obtained by:

\begin{equation}
    D^{*}\left(x\right) = \frac{p_{r}\left(x\right)}{{p_{r}\left(x\right)}+ {p_{g}\left(x\right)}}
\end{equation} 
In the course of employment, only $G$ is utilized to produce new synthetic data while $D$ is usually discarded.

\subsection{Model extraction attacks against machine learning models}
A machine learning model is essentially a function ${f}$ that maps input data $\mathcal{X}$ to output data $\mathcal{Y}$: $\mathcal{Y} = {f}(\mathcal{X})$. In general, machine learning models can be categorized as two classes~\cite{StatisticalMechanics}: discriminative models and generative models. For discriminative models on image classification tasks, the input data corresponds to an image while the output data can be interpreted as a probability distribution over categorical labels. \textit{A key goal of discriminative models} is to find an optimal set of parameters which minimizes the errors on the test dataset. For generative models on image generation tasks, the input data is represented by a latent code and the output data is an image. \textit{A core goal of generative models} is to adjust the parameters to learn a distribution which is similar to the training data distribution $p_{r}$.
 
A model extraction attack in the machine learning setting emerges when an adversary aims to duplicate a model $\tilde{f}$ through querying the target model $f$. In general, there are two types of attacks around model extraction based on adversary's objective: accuracy extraction and fidelity extraction~\cite{jagielski2019high}. For discriminative models, accuracy extraction requires the extracted model to match or exceed the accuracy of the target model on the test 
dataset, while fidelity extraction requires the extracted model not only to achieve the same accuracy as the target model on the test dataset but also to replicate the errors of the target model. The limit of fidelity extraction is the functionally-equivalent model extraction~\cite{jagielski2019high}. Considering different goals and evaluations between discriminative models and generative models, we redefine model extraction on GANs in Section~\ref{ssec:adversarygoals}.

\subsection{Dataset description}
We utilize five different datasets in this paper, which are all widely adopted in image generation. Among them, four datasets are from the LSUN dataset~\cite{yu2015lsun} which includes 10 scene categories and 20 object categories and we define them as LSUN-Bedroom, LSUN-Church, LSUN-Classroom, and LSUN-Kitchen, respectively. CelebA dataset~\cite{liu2015faceattributes} consists of about 200k high-quality human face images. Datasets including LSUN-Bedroom, LSUN-Classroom, and LSUN-Kitchen are only used in Section \ref{sec:casestudy1} to illustrate the attack effects in a case study. The details of the datasets are shown in Table~\ref{tab:Dataset description}.

\begin{table}[!t]
	\centering	
	\caption{Dataset description}
	\label{tab:Dataset description}
	\begin{tabular}{lrrr}
		\toprule
		Dataset & LSUN-Bedroom & LSUN-Kitchen & CelebA \\
		Size of dataset & 3,033,042 &  2,212,277 & 202,599\\
		\midrule
		Dataset & LSUN-Classroom  & LSUN-Church\\
		Size of dataset & 168,103 &126,277\\
		\bottomrule
	\end{tabular}
\end{table}
  

%% file: 4_Taxonomy.tex
\section{Taxonomy of model extraction against GANs}  
\label{sec:taxonomy}
In this section, we start with adversary's goal and formally elaborate on our attacks. Next, we illustrate adversary's background knowledge where an adversary can mount attacks according to the obtained information. Finally, we detail the metrics to evaluate the performance of attack models.
\begin{table}[!t]
\centering
	\caption{Notations}
	\label{tab:Notations}
	\begin{tabular}{ll}
		\toprule
		Notation & Description\\
		\midrule
		{$p_{r}$} & distribution of training set of a GAN\\
		{$p_{g}$} & implicit distribution of a target generator\\
		{$\Tilde{p}_g$} & implicit distribution of an attack generator\\
		{$\it accuracy$} & FID ($\Tilde{p}_g$, $p_g$)\\
		{$\it fidelity$} & FID ($\Tilde{p}_g$, $p_r$)\\		
		\bottomrule
	\end{tabular}
\end{table}

\subsection{Adversary's goals}
\label{ssec:adversarygoals}
%
In general, model extraction based on adversary's goals can be categorized into either accuracy extraction or fidelity extraction. Unlike supervised discriminative models aiming at minimizing errors on a test set, unsupervised generative models target at learning the distribution of a data set. Therefore, for model extraction attacks on GANs, accuracy extraction aims to minimize the difference of data distribution between attack models and target models, while fidelity extraction aims to minimize the distribution between attack models and the training set of target models. 

\begin{figure*}[!t]
	\centering
	\includegraphics[width=0.90\linewidth]{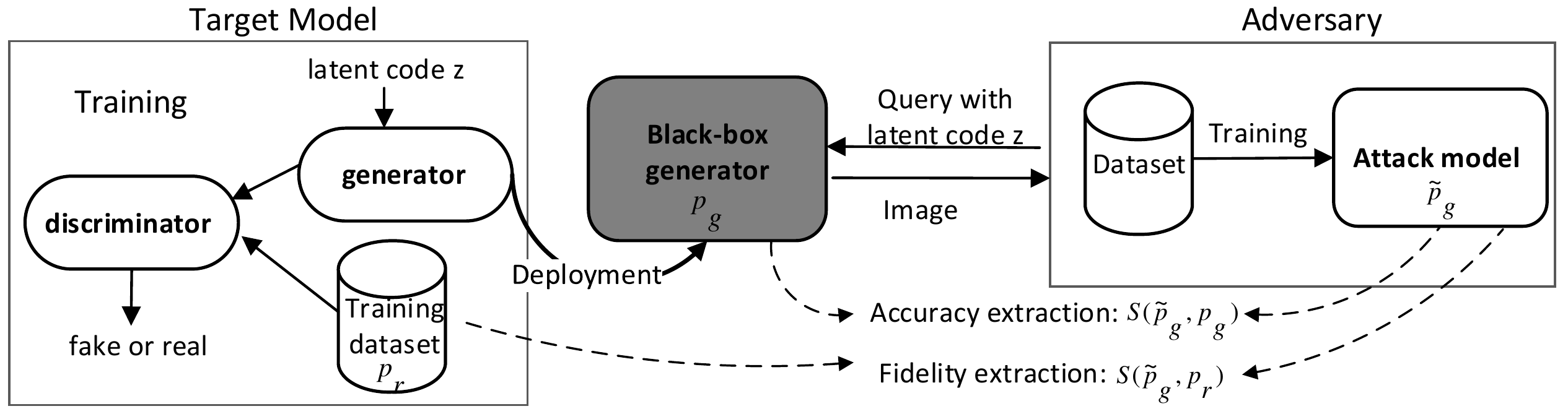}
	\caption{Attack types: accuracy extraction and fidelity extraction.}
	\label{fig:attack types}
\end{figure*}

Specifically, as shown in Figure~\ref{fig:attack types}, the goal of accuracy extraction is to construct a $\Tilde{G}$ minimizing $ S(\Tilde{p}_{g},p_g) $, where $S$ is a similarity function, $\Tilde{p}_{g}$ is the implicit distribution of the attack generator $\Tilde{G}$, and $p_g$ is the implicit distribution of the target generator $G$. In contrast, fidelity extraction's goal is to construct a $\Tilde{G}$ minimizing $ S( \Tilde {p}_{g},p_r) $, where $p_r$ is the distribution of the training set of the target generator $G$. 
In this work, we use Fr{\'e}chet Inception Distance~(FID) to evaluate the similarity between two data distributions, mainly considering its computational efficiency and robustness~\cite{FIDl2017gans}. It is elaborated in Section~\ref{ssec:metrics}. In our work, we study the accuracy extraction in Section~\ref{sec:accuracy}, and fidelity extraction in Section~\ref{sec:fidelity}. 
 
\subsection{Adversary's background knowledge}
\begin{figure}[!ht]
	\centering
	\includegraphics[width=1.0\linewidth]{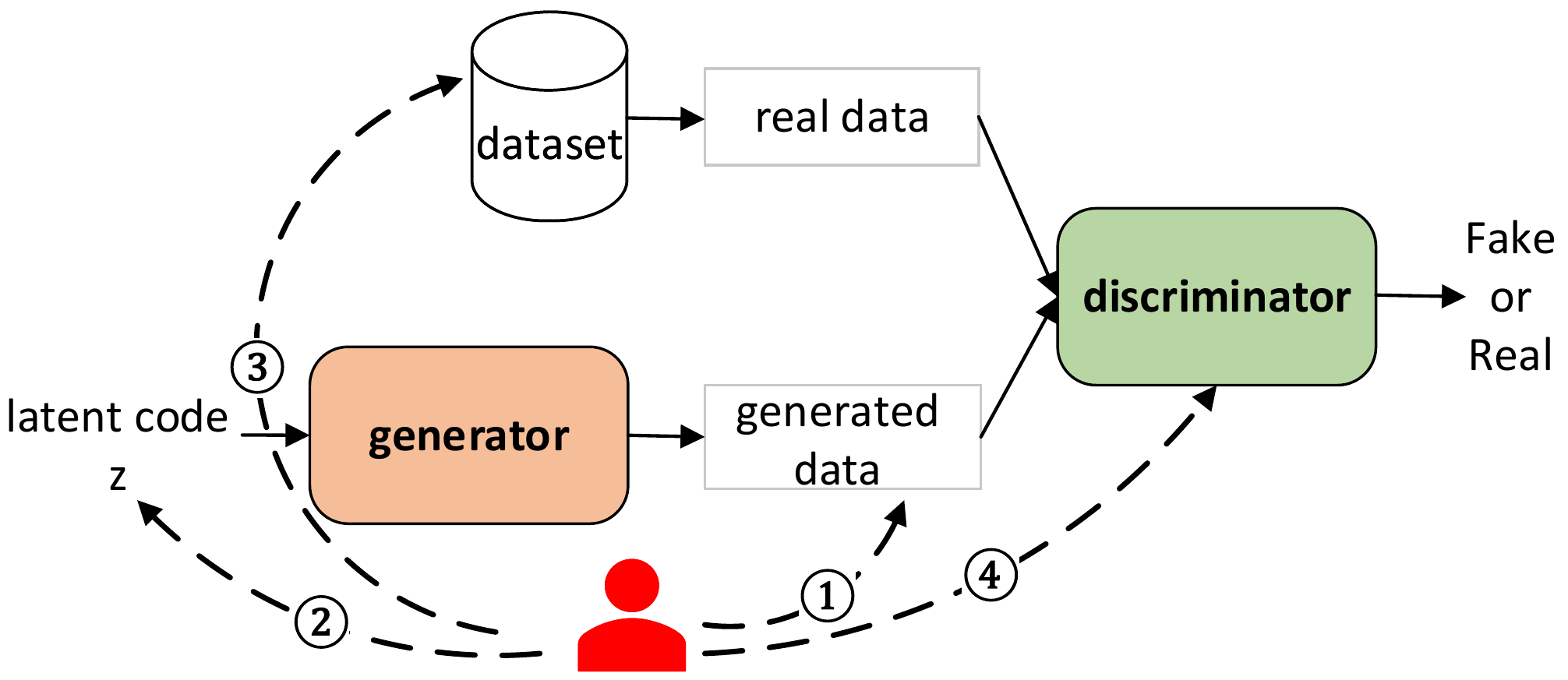}
	\caption{Adversary's background knowledge. }
	\label{fig:Attack settings}
\end{figure}

Adversaries can mount model extraction attacks at different levels based on their obtained information about the target GAN. The more background knowledge adversaries acquire, the more effective they should be in achieving their goal. In general, four components of a GAN can be considered by an adversary. As shown in Figure \ref{fig:Attack settings}, they are respectively: (1)~generated data; (2)~latent codes used by interactively querying a generator; (3)~partial real data from the training dataset of the target GAN; (4)~a discriminator from the target GAN.
\begin{enumerate}
	\item[$\bullet$] \textit{Generated data:} generated data refers to samples from the generator of the target GAN. In some scenarios, adversaries are unable to query the target model but only have access to the generated samples which are publicly released by the model provider. Whether they can steal a good-quality model largely depends on the amount of released data from model owners.
	\item[$\bullet$] \textit{Latent codes:} latent codes refer to random numbers drawn from a prior distribution such as Gaussian distribution. Adversaries obtaining latent codes mean that they know the prior distribution and use it to interactively query the target model's API provided by model owners. This is a basic assumption in the current query-based attack setting.
	\item[$\bullet$] \textit{Partial real data:} this is more knowledgeable for adversaries when they can get partial real data which is used to train the target GAN. But this is also a common assumption which can be found in many literature in relation to security and privacy of machine learning \cite{ji2015your,shokri2017membership,hayes2019logan}.
	\item[$\bullet$] \textit{Discriminator:} adversaries may obtain the discriminator of the target GAN although it is generally removed in the deployment. In our work, we illustrate how the discriminator can be explored to improve the $\it fidelity$ of attack models.
\end{enumerate}

In the following attack settings, we assume an adversary obtains different levels of background knowledge to achieve accuracy extraction or fidelity extraction.

\subsection{Metrics}
\label{ssec:metrics}
In this work, we evaluate the performance of GANs by the FID~\cite{FIDl2017gans} which measures the similarity between $p_g$ and $p_r$. Specifically, on the basis of features extracted by the pretrained Inception network $\phi$, it models $\phi(p_r)$ and $\phi(p_g)$ using Gaussian distribution with mean $\mu$ and covariance $\Sigma$, and the value of FID between real data $p_r$ and generated data $p_g$ in convolutional features is computed as: ${\it FID}(p_r, p_g) = \left | \left | \mu _r -\mu _g \right | \right | + Tr(\Sigma _r + \Sigma _g -2(\Sigma _r \Sigma _g)^{1/2})$. A lower FID indicates that the distribution's discrepancy between the generated data and real-world data is smaller and the generated data is more realistic. In our work, FID is computed  with all real samples and 50k generated samples.
 
In addition to FID which evaluates the performance of a model itself, we also need indicators to measure whether an attack is considered to be successful. In this work, we use two FID-based metrics: {\it accuracy} and {\it fidelity}, to evaluate the performance of attack models. {\it Accuracy} measures the consistency between $p_g$ which is an implicit distribution of a target generator and $\Tilde{p}_g$ which is an implicit distribution of an attack generator. Note that, $\it accuracy$ not only measures how close the attack model and the target model are, but also indicates how well the performance of model itself is.
In contrast, {\it fidelity} measures the consistency of data distribution between $p_r$ and $\Tilde{p}_g$. 
Similar to FID, the smaller the {\it accuracy} and {\it fidelity} values are, the better performance attack models achieve. When it is clear from the context, we refer to {\it accuracy} and {\it fidelity} as {\it accuracy} value and {\it fidelity} value, respectively. The summarized notations can be seen in Table~\ref{tab:Notations}. 

Accuracy extraction focuses on {\it accuracy} and adversaries aim to steal the distribution of a target model. After obtaining an attack model which steals from a target model, they can directly utilize it to generate new samples. Additionally, they can also transfer knowledge of the stolen model to their own domains through transfer learning. In contrast, fidelity extraction concentrates on {\it fidelity} and adversaries target at stealing the distribution of the training set of a target model. This type of attacks can severely violate the privacy of the training data and it also means that adversaries may steal valuable commercial datasets from a trained GAN. Additionally, adversaries can utilize the stolen high-fidelity model to mount other novel attacks and we leave it for future work.

%% file: 5_accuracy.tex
\section{Accuracy extraction}
\label{sec:accuracy}
In this section, we instantiate our accuracy extraction attack strategy. we assume that adversaries have access to either generated samples provided by the model producer or querying the target model to obtain data. Therefore, whether only publicly access to data generated by GANs or black-box querying can leak the information of models is a question. In this section, we start with target models and attack models. Then, we describe our attack performance. Next, we study the effect of the number of queries and queries from different prior distributions. In the end, we perform experiments to deeply understand model extraction on GANs.

\subsection{Target models and attack models}
\label{ssec:Target models and attack models}
We choose representative GANs: Progressive GAN (PGGAN)~\cite{PGGAN2018progressive} and Spectral Normalization GAN (SNGAN)~\cite{SNGAN2018spectral} as our target models, which both show pleasing performances in image generation.
The implementation details can be seen in Appendix~\ref{ssec:Implementation details}. For training sets LSUN-Church and CelebA, we first resize them to 64 $\times$ 64 and use all records of each dataset to train our target models. As shown in Table~\ref{tab:target GAN's performance}, target GAN models achieve an excellent performance on these dataset and the performance of PGGAN is better than that of SNGAN. 

We use GANs as our attack models to extract target models. In practice, adversaries may not know the target model's architecture. Therefore, we study the performance of attack models with different architectures. Specifically, we choose SNGAN and PGGAN as our attack models. There are four different situations for their combinations. For simplification, we define each situation as an attack-target model pair, and they are respectively SNGAN-SNGAN, SNGAN-PGGAN, PGGAN-SNGAN and PGGAN-PGGAN. The reason why we choose SNGAN and PGGAN as the research object is that: 1) they both show good performance in image generation;  and 2) they have significant difference in the aspects of training, loss function and normalization, which all facilitate us to study the performance of attack models with different architectures.  

\begin{table}[!t]
	\centering
	\caption{Performance of target GANs (see Figure~\ref{fig:target_gan} in Appendix for qualitative results).}	
	\label{tab:target GAN's performance}
	\begin{tabular}{llr}
		\toprule
		Target model & Dataset & FID \\
		\midrule
		SNGAN & LSUN-Church &12.72 \\
		SNGAN & CelebA & 7.60 \\
		PGGAN & LSUN-Church & 5.88 \\
		PGGAN & CelebA & 3.40 \\
		\bottomrule
	\end{tabular}
\end{table}   

\subsection{Methodology}
\label{ssec:Accuracy Methodology}
As shown in Figure~\ref{fig:attack types}, for accuracy extraction, we assume that an adversary obtains the generated data by the model provider or querying the target GAN. This scenario is practical, because some model owners need to protect their models through providing the public with some generated data or a black-box GAN model API.
In this case, the adversary uses the generated data to retrain a GAN to extract the target model.
We do not distinguish whether generated data is from queries or model providers, because our approach only relies on these generated data. However, in Section~\ref{sssec:different prior distributions}, we study attack performance on queries with different prior distributions.

Note that model extraction on GANs is different from machine learning on GANs. This is because machine learning on GANs requires users to train a GAN on real samples which are collected from the real world. In contrast, model extraction on GANs enables users to train a GAN on generated data from a target GAN model. In essence, model extraction on GANs approximates the target GAN which is a much simpler deterministic function, compared to real samples which usually represents a more complicated function. 

\subsection{Results}
\label{ssec:Accuracy results}
\subsubsection{\textbf{Attack performance on different models.}}
Table~\ref{tab:accuracy attack} shows the accuracy extraction's performance with 50k queries to the target model. In general, attack models can achieve an acceptable performance. For instance, our attack performance of PGGAN-PGGAN on the CelebA achieves 1.02 FID on {\it accuracy}, which means that the attack model can achieve a perfect extraction attack for the target model.\footnote{Although we cannot directly compare the performance of different models on different datasets, we can choose the state-of-the-art StyleGAN~\cite{StyleGAN12019style} trained on the LSUN-Bedroom dataset as a reference, where it has the lowest FID 2.65.} 
It is noticeable that the the attack model achieves such performance only on 50k generated images while the target model is trained on more than 200k images, which indicates that the privacy of the target model can be leaked by their generated images. In other words, adversaries are able to obtain a good GAN model only by access to the generated data from the target model instead of collecting their own data which is usually labor-intensive and time-consuming. 

For the target model PGGAN, if the attack model is SNGAN, we observe that the performance of model extraction is very efficient on both CelebA and LSUN-Church dataset and the attack model SNGAN can learn more from the target model PGGAN, compared to the SNGAN-SNGAN case, which indicates that attacking a state-of-the-art GAN is valuable and viable for an adversary. Furthermore, this case SNGAN-PGGAN is the most common situation in the actual attack scenarios, because generally we implicitly assume that performance of the adversary's model may often be weaker than that of the target model and the structure of the attack model is inconsistent with that of the target model. 

We also report {\it fidelity} in Table~\ref{tab:accuracy attack} and find that for model extraction on GAN models, the  {\it fidelity} of attack models is always higher than that of target model, in which  {\it fidelity} of attack models represents similarity between distribution of real dataset $p_r$ and distribution of the attack model $\Tilde{p}_g$ and for  {\it fidelity} of a target model, also called FID of target model, it represents similarity between distribution of real dataset $p_r$ and distribution of the target model ${p}_g$. For example, when the target model SNGAN has 12.72 FID on the LSUN-Church dataset,  {\it fidelity} of the attack model SNGAN will increase to 30.04. Even for the PGGAN-PGGAN case, its  {\it fidelity} increases from 3.40 to 4.93 on the CelebA dataset. This is mainly because although theoretically, the distribution of the target model $p_g$ is equal to that of the real training dataset $p_r$,  it is actually not equal because GAN cannot achieve the global optimum. However, we will discuss how to reduce  {\it fidelity} and achieve fidelity extraction in Section~\ref{sec:fidelity}.      

For the target model SNGAN, if the attack model is PGGAN, the {\it accuracy} of model extraction is lower than that of the attack model SNGAN. It is mainly because the PGGAN model itself is stronger and able to more accurately approximate the target model. Similarly, PGGAN as an attack model has more lower {\it fidelity}, in contrast with SNGAN as an attack model. For instance, compared to SNGAN-SNGAN with 17.32 of {\it fidelity} on CelebA dataset, the {\it fidelity} of PGGAN-SNGAN is only 9.57, which largely improves the attack performance on fidelity. This indicates that using an attack model which is larger than the target model is an efficient approach to improve attack performance. 

Overall, accuracy extraction can achieve a good performance in terms of {\it accuracy}. In general, adversaries can steal an accurate model, and then use the extracted model for their own purpose. However, unlike discriminative models where adversaries can directly utilize their extracted model, the extracted model of a GAN only generates target model's images. Therefore, in Section~\ref{sec:casestudy1}, we will perform a case study where adversaries can effectively leverage the extracted model to generate images for their own applications rather than target GANs' images through transfer learning.

\begin{table}[!t]
\centering
	\caption{The performance of accuracy extraction with 50k queries to the target model (see Figure~\ref{fig:target_pggan} in Appendix for qualitative results).}	
	\label{tab:accuracy attack}
	\renewcommand{\arraystretch}{1.3}
	\scalebox{0.75}{
	\begin{tabular}{lllrrr}
		\toprule
		Target model & Attack model & Dataset & {\it Accuracy} & {\it Fidelity} \\
		 &  &  & FID($\Tilde{p}_g$, $p_g$) & FID ($\Tilde{p}_g$, $p_r$)\\
		\midrule

		\multirow{4}{*}{PGGAN} &SNGAN & LSUN-Church &6.11 & 14.05\\
		&SNGAN & CelebA &4.49 & 9.29\\
		&PGGAN & LSUN-Church &1.68 & 8.28\\
		&PGGAN & CelebA &1.02 & 4.93\\
				\cline{1-5}
		\multirow{4}{*}{SNGAN} &SNGAN & LSUN-Church &8.76 & 30.04\\
		&SNGAN & CelebA &5.34 & 17.32\\		
		&PGGAN & LSUN-Church &2.21 & 14.56\\
		&PGGAN & CelebA &1.39 & 9.57\\

		\bottomrule
	\end{tabular}}
\end{table}

\subsubsection{\textbf{Attack performance on the number of queries.}}
\label{ssec:Accuracy different queries}
\begin{figure}[!t]
\centering
  \subfigure[{\it Accuracy} on CelebA]{
	\includegraphics[width=0.450\columnwidth]{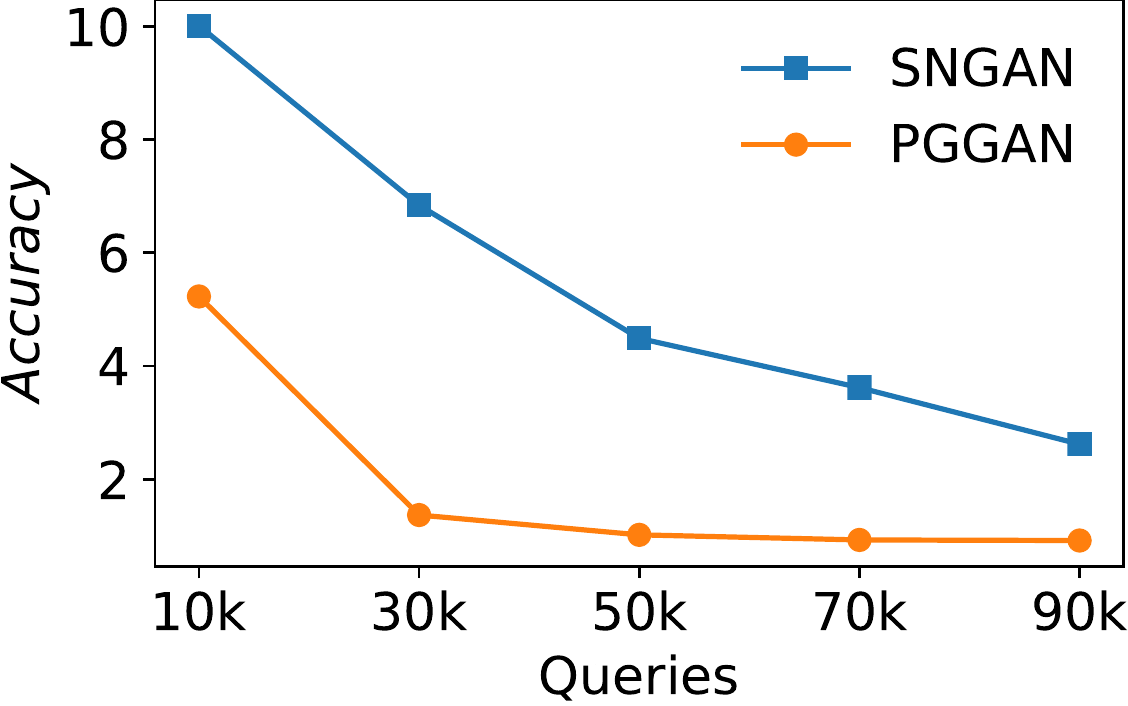}
	\label{fig:accuracy_different_sizes}
    }
  \subfigure[{\it Fidelity} on CelebA]{
	\includegraphics[width=0.450\columnwidth]{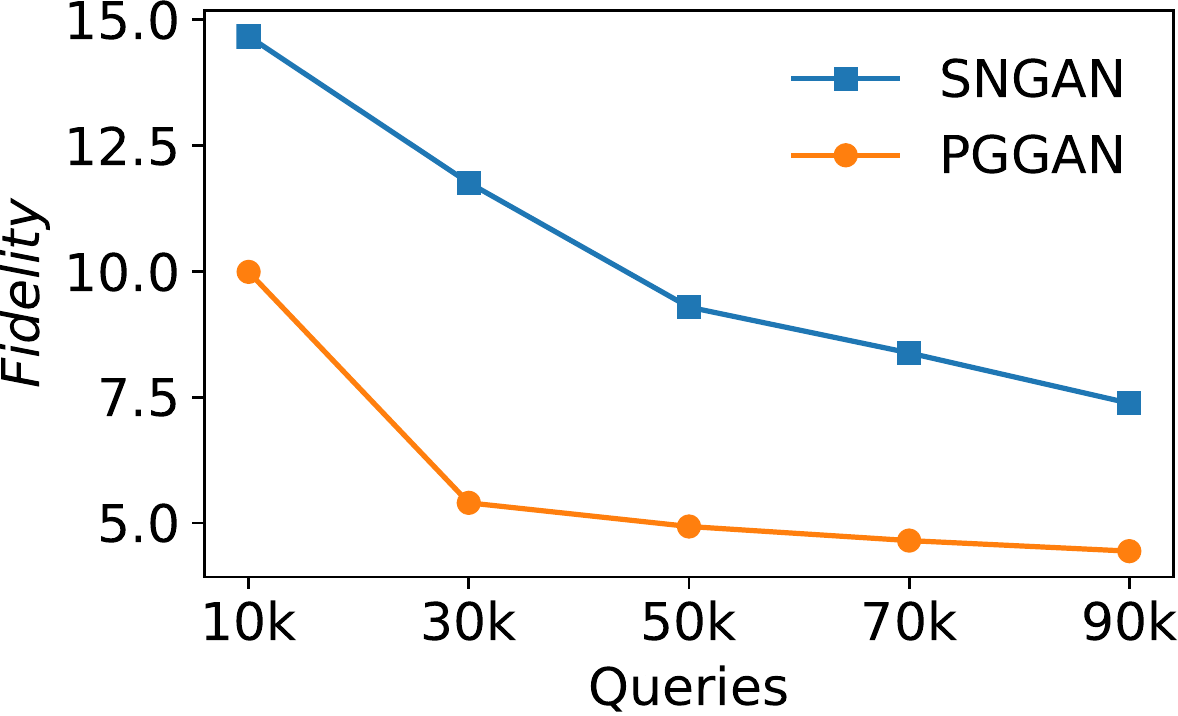}
	\label{fig:fidelity_different_sizes}
    }
    \caption{Attack performance on the number of queries.}
    \label{fig:different_sizes}
\end{figure}

We choose PGGAN trained on CelebA dataset as the target model to study the effect of the number of queries due to the best performance among our target models. Figure~\ref{fig:different_sizes} plots the attack performance with respect to the number of queries which are also the size of training dataset of attack models. As expected, we observe that the attack performance increases with an increase in queries. For instance, when the number of queries increases to 90k, attack models' {\it fidelity} and {\it accuracy} are both low. However, as the number of queries decreases, the attacks become less effective due to the lack of sufficient training data for the attack models. Especially for the attack model SNGAN with 10k queries, its {\it fidelity} and {\it accuracy}  are both over 10.00. This indicates that releasing a small number of data by the model owner or restricting the number of queries is a relatively safe measure, because it is difficult for an adversary to extract a model with limited data.

\subsubsection{\textbf{Attack performance on queries from different prior distributions.}} 
\label{sssec:different prior distributions}
Adversaries can query the target model via trying common prior distributions to generate latent codes if they do not know the prior distribution of a target model. Gaussian distribution and uniform distribution are widely used in almost all GANs~\cite{PGGAN2018progressive,BigGAN2018large,DCGAN2015unsupervised,SNGAN2018spectral,StyleGAN12019style,StyleGAN22019analyzing}. Table~\ref{tab:prior distribution} shows the attack performance with two prior distributions. We choose PGGAN trained on CelebA dataset with standard normal prior distribution as the target model. From Table~\ref{tab:prior distribution}, we find that adversaries can obtain a similar attack performance no matter what the prior distribution of latent codes is.

\begin{table}[!t]
\centering
	\caption{Performance of accuracy extraction attack with different prior distributions. We use standard normal distribution and uniform distribution over an interval -1 and 1 to generate latent codes. The number of queries is fixed to 50k.}	
	\label{tab:prior distribution}
	\renewcommand{\arraystretch}{1.1}
	\scalebox{0.95}{
	\begin{tabular}{llrr}
		\toprule
		Attack model & Prior distribution & {\it Accuracy} & {\it Fidelity} \\
		 &  &FID($\Tilde{p}_g$, $p_g$) & FID ($\Tilde{p}_g$, $p_r$)\\
		\midrule

		SNGAN& Gaussian &4.49 & 9.29\\
		SNGAN& Uniform &4.29 & 9.16\\
		PGGAN& Gaussian &1.02 & 4.93\\
		PGGAN& Uniform 
		&0.98 &4.85\\

		\bottomrule
	\end{tabular}}
\end{table}

\subsubsection{\textbf{Understanding accuracy extraction on GANs in-depth.}} 
\label{Understanding accuracy extraction}
\begin{figure}[!t]
\centering
	\includegraphics[width=1\columnwidth]{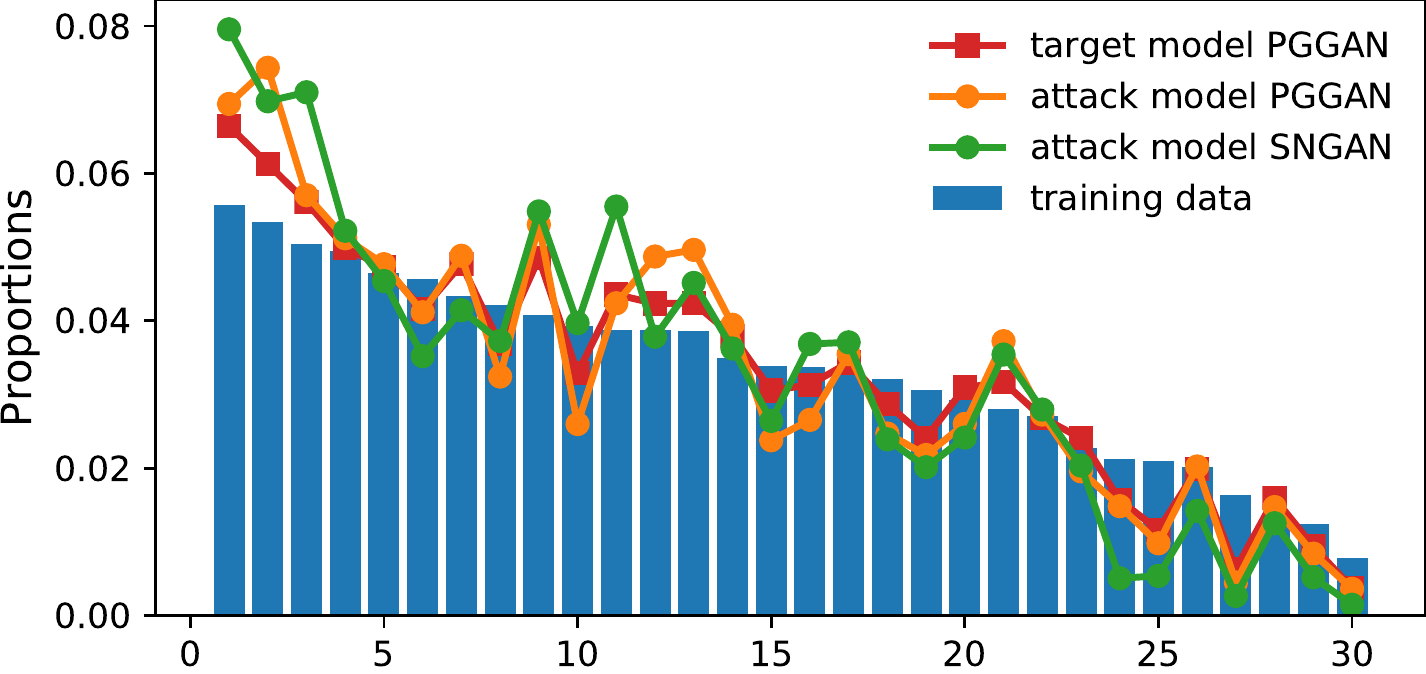}	
    \caption{Class distributions of the training data, the target model PGGAN, and attack models.}
    \label{fig:dissect_accuracy_pggan}
\end{figure}

We further dissect the difference of distributions between target models and attack models to understand the nature of model extraction on GANs.
Specifically, we first transform the training data into 2048-dimension feature vectors by the pretrained Inception-v3 model which is widely utilized in the evaluation of a GAN model. Then these feature vectors are clustered into $k$ classes by a standard $K$-means algorithm. Finally, we calculate the proportions of each class and proportions of all classes can be considered as a distribution of the training data~\cite{FSD2019seeing,richardson2018gans}. 
The blue bar in Figure~\ref{fig:dissect_accuracy_pggan} shows the distribution of the training data where we set $k$ to 30. 
For target models and attack models, we query the model to obtain 50k images, then perform the same procedures as the training data. 

Figure~\ref{fig:dissect_accuracy_pggan} shows distribution differences among the training data, the target model PGGAN and attack models. 
We observe that for the high proportions of classes, which can be considered as prominent features of a distribution, target models can learn more features about these classes while attack models further learn more features by querying the target models. In contrast, for the low proportions of classes, target models learn less features about these classes while attack models further learn less features about these classes. 
This is one reason why attack models always have higher {\it fidelity} than target models. In terms of {\it accuracy}, we observe that there is a consistent trend on proportions of classes for target models and attacks models. This is the reason why we can achieve a good performance about {\it accuracy}. We also analyze the target model SNGAN, and similar results are shown in Figure~\ref{fig:dissect_accuracy_sngan} in Appendix. 

We also summarize this difference in a single number by computing the Jensen-Shannon~(JS) divergence on this representation of distributions, which is shown in Table~\ref{tab:accuracy attack_js}. Note that, based on {\it accuracy} and {\it fidelity} defined in Section~\ref{ssec:adversarygoals}, we mark $JS_{\it accuracy}$ as the JS divergence between the target model and the attack model, and $JS_{\it fidelity}$ as the JS divergence between the training data and the attack model.

\begin{table}[!t]
\centering
	\caption{JS distances between models. A smaller value indicates a better performance. For the JS distance between training data and the target model, the target model PGGAN is $4.14 \times 10^{-3}$. The JS value shows a consistent trend with Figure~\ref{fig:dissect_accuracy_pggan}. }	
	\label{tab:accuracy attack_js}
	\renewcommand{\arraystretch}{1.1}
	\scalebox{0.9}{
	\begin{tabular}{llrr}
		\toprule
		Target model & Attack model & $\it JS_{\it accuracy}$ ($\times 10^{-3}$)& $\it JS_{\it fidelity}$ ($\times 10^{-3}$)\\
		\midrule

		\multirow{2}{*}{PGGAN} &SNGAN  &5.88 & 15.95\\
		&PGGAN & 1.83 & 9.10\\
		\bottomrule
	\end{tabular}}
\end{table}

%% file: 6_fidelity.tex
\section{Fidelity extraction}
\label{sec:fidelity}
In this section, we instantiate our fidelity extraction attack strategy. In addition to accuracy extraction's assumptions, we also assume that adversaries have more background knowledge in order to achieve fidelity extraction, such as partial real data or target model's discriminator. We start with the motivation and problem formulation of fidelity extraction. Then, we describe the methodology of fidelity extraction. In the end, we present the performance of fidelity extraction.  

\subsection{Motivation and problem formulation}
As shown in Figure~\ref{fig:attack types}, accuracy extraction can be implemented through querying the generator of the target GAN, because $p_g$ is the generator's distribution. As for fidelity extraction, it is much more difficult due to the lack of availability of real data distribution $p_r$. Although an approach is to use $p_g$ as an approximation of $p_r$, we observe that with the increase in the number of queries, {\it fidelity} of attack models reaches its saturation point and is hard to be improved, which is shown in Figure~\ref{fig:fidelity_different_sizes}. For instance, as we increase the number of queries from 50k to 90k for the PGGAN-PGGAN case on CelebA dataset, {\it fidelity} of the attack model has smaller and smaller improvements from 4.93 to 4.44, while the ideal {\it fidelity} is 3.40 which is also the performance of the target model. Note that the case PGGAN-PGGAN is the best for the attacker; the attack will perform even worse if the attackers do not choose the same architectures and hyperparameters as the target model.  

\begin{figure*}[!t]
\centering
  \subfigure[Real Samples 50k]{
    \includegraphics[width=0.35\columnwidth]{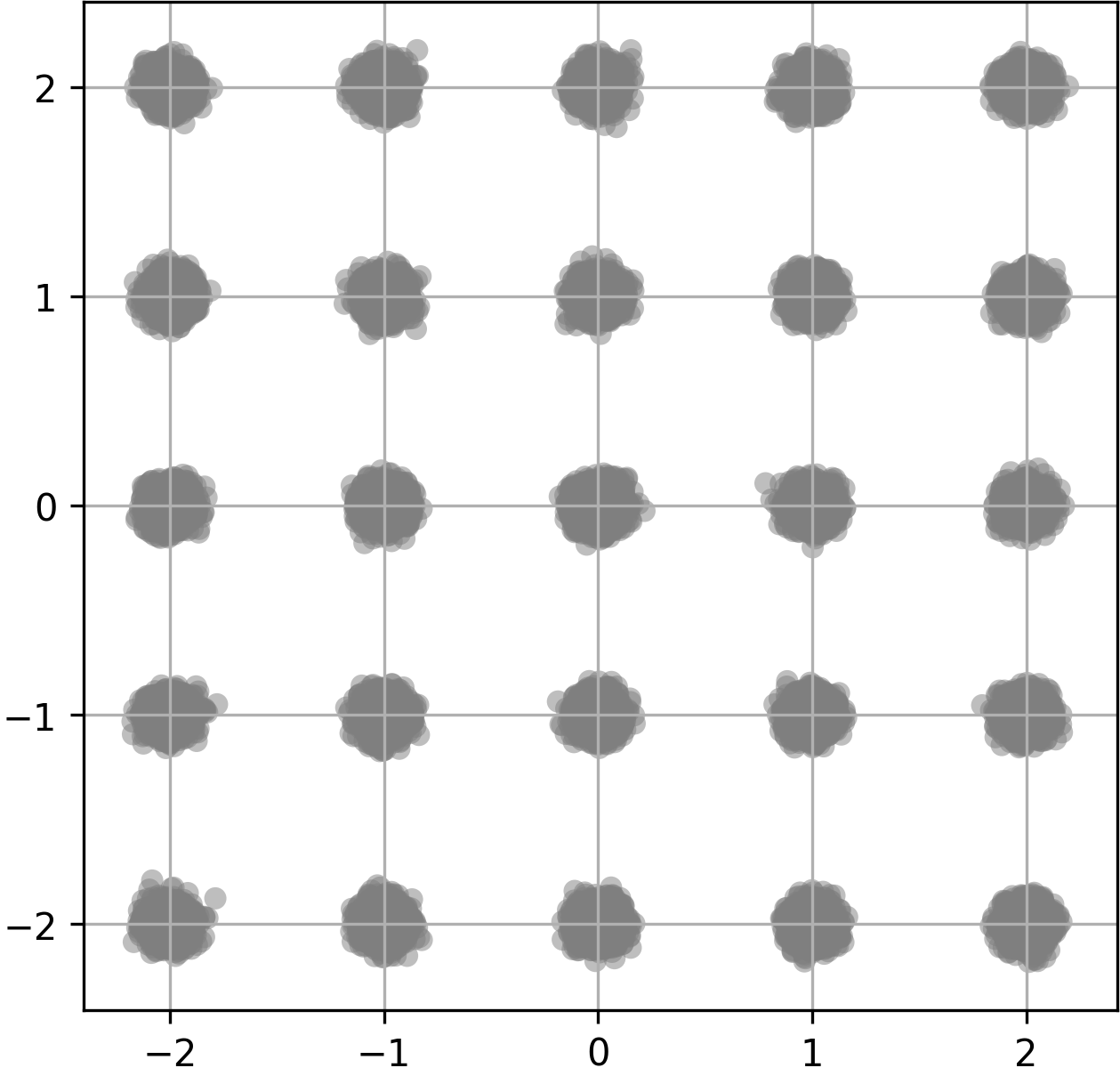}
    \label{fig:real_samples}
    }
  \subfigure[Direct Sampling 5k]{
    \includegraphics[width=0.35\columnwidth]{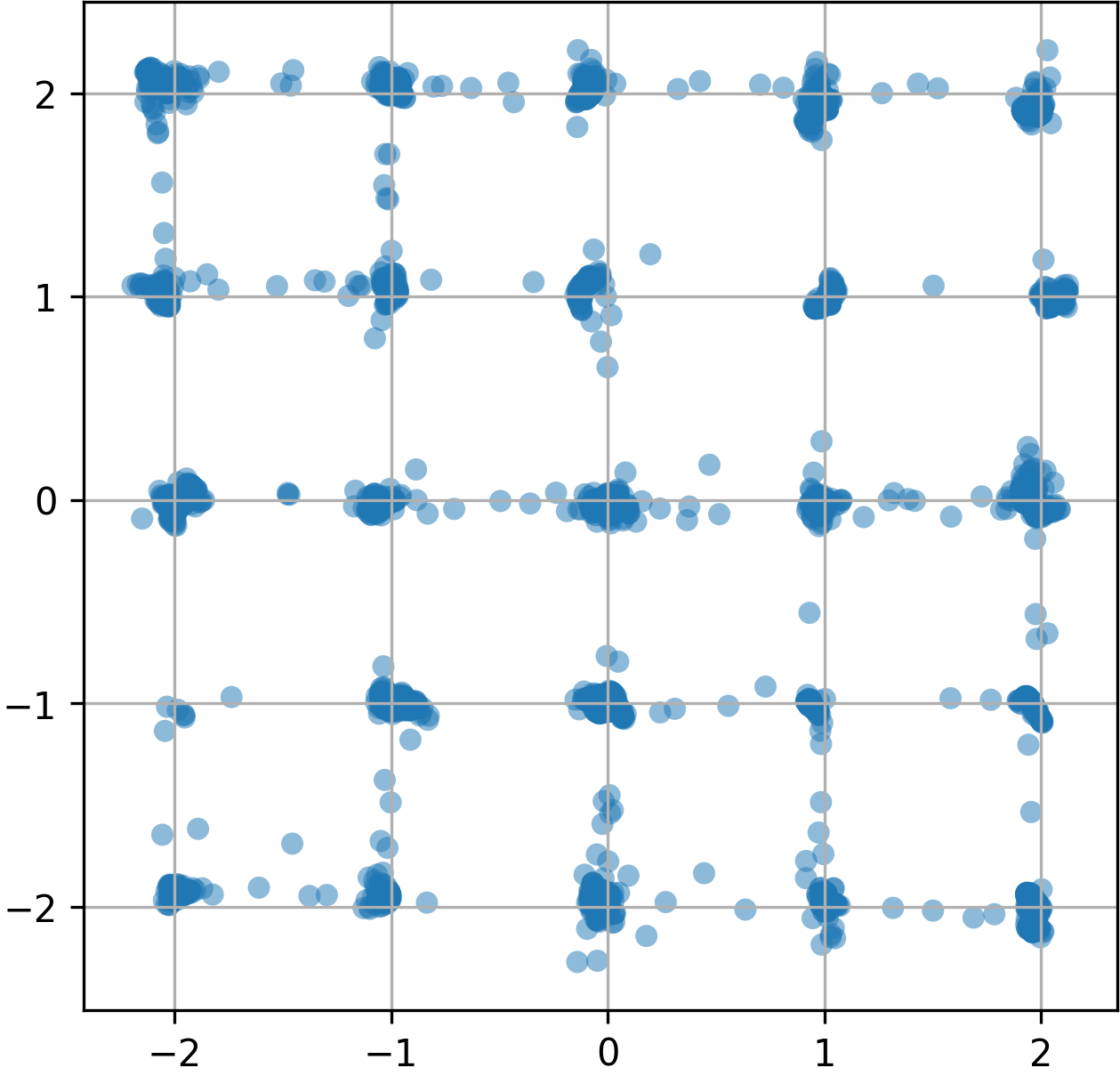}
    \label{fig:direct_samples_5k}
    }
  \subfigure[Direct Sampling 10k]{
	\includegraphics[width=0.35\columnwidth]{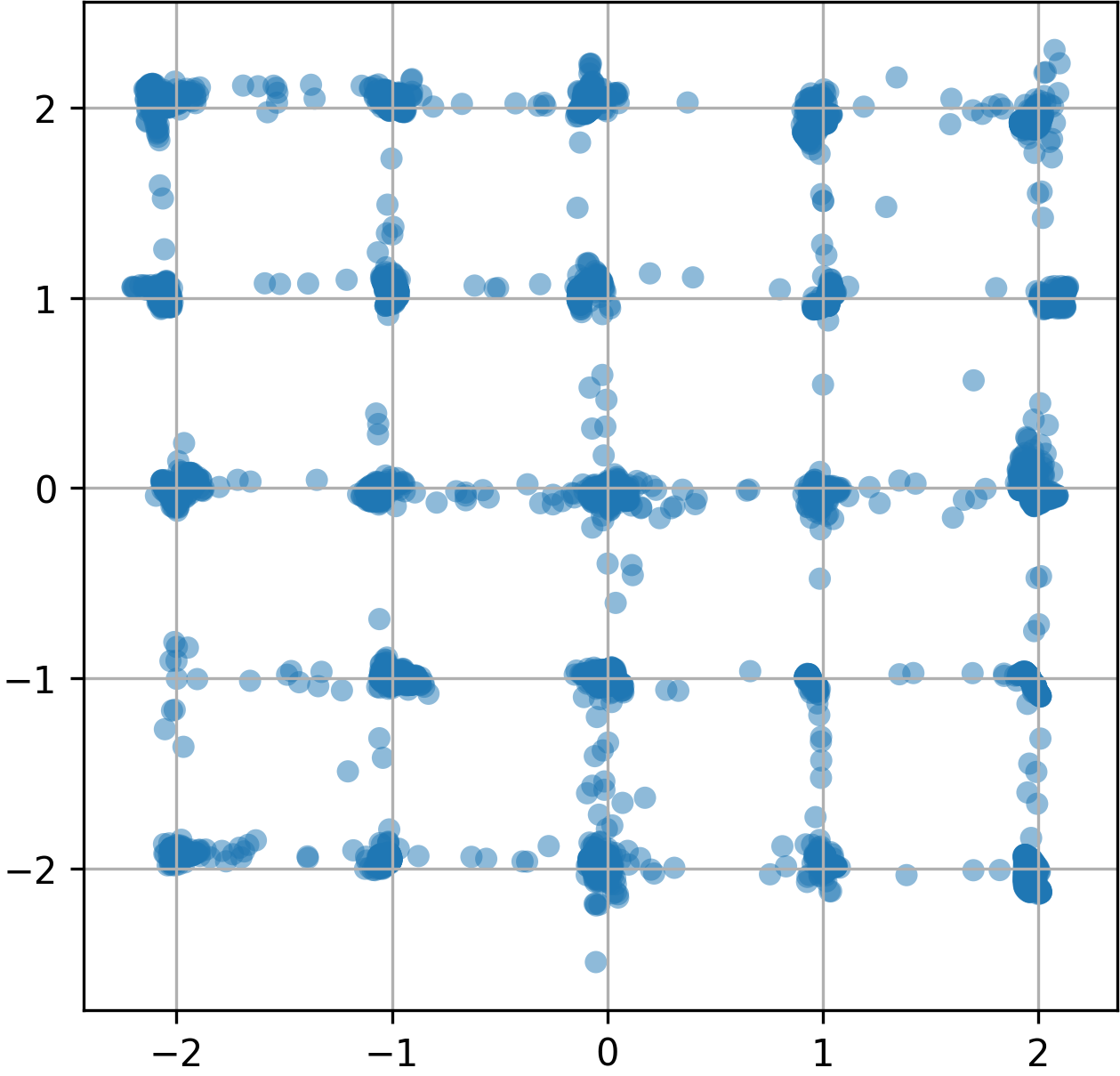}
	\label{fig:direct_samples_10k}
	}
  \subfigure[Direct Sampling 50k]{
    \includegraphics[width=0.35\columnwidth]{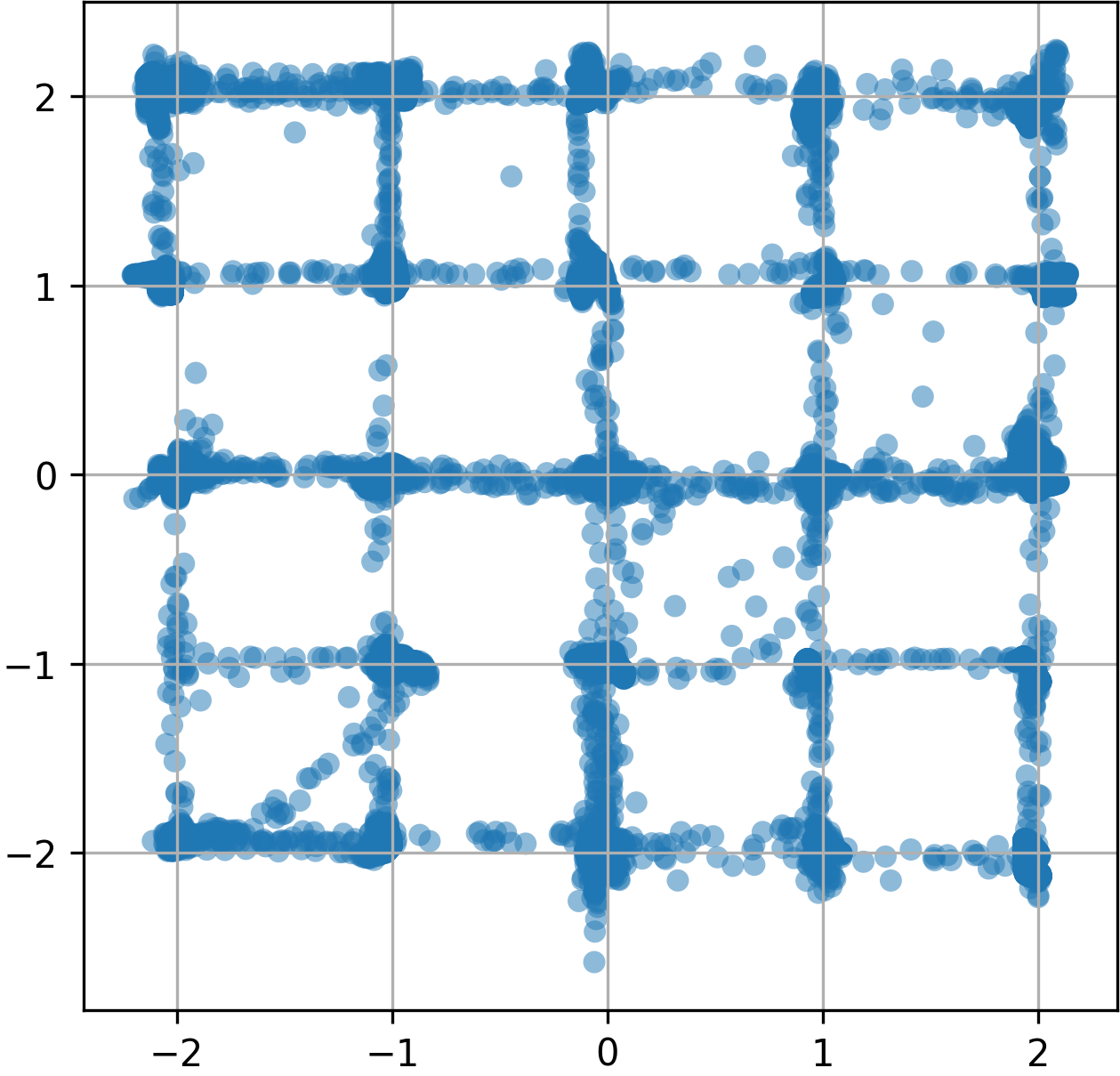}
    \label{fig:direct_samples_50k}
    }
    \subfigure[MH Subsampling 50k]{
    \includegraphics[width=0.35\columnwidth]{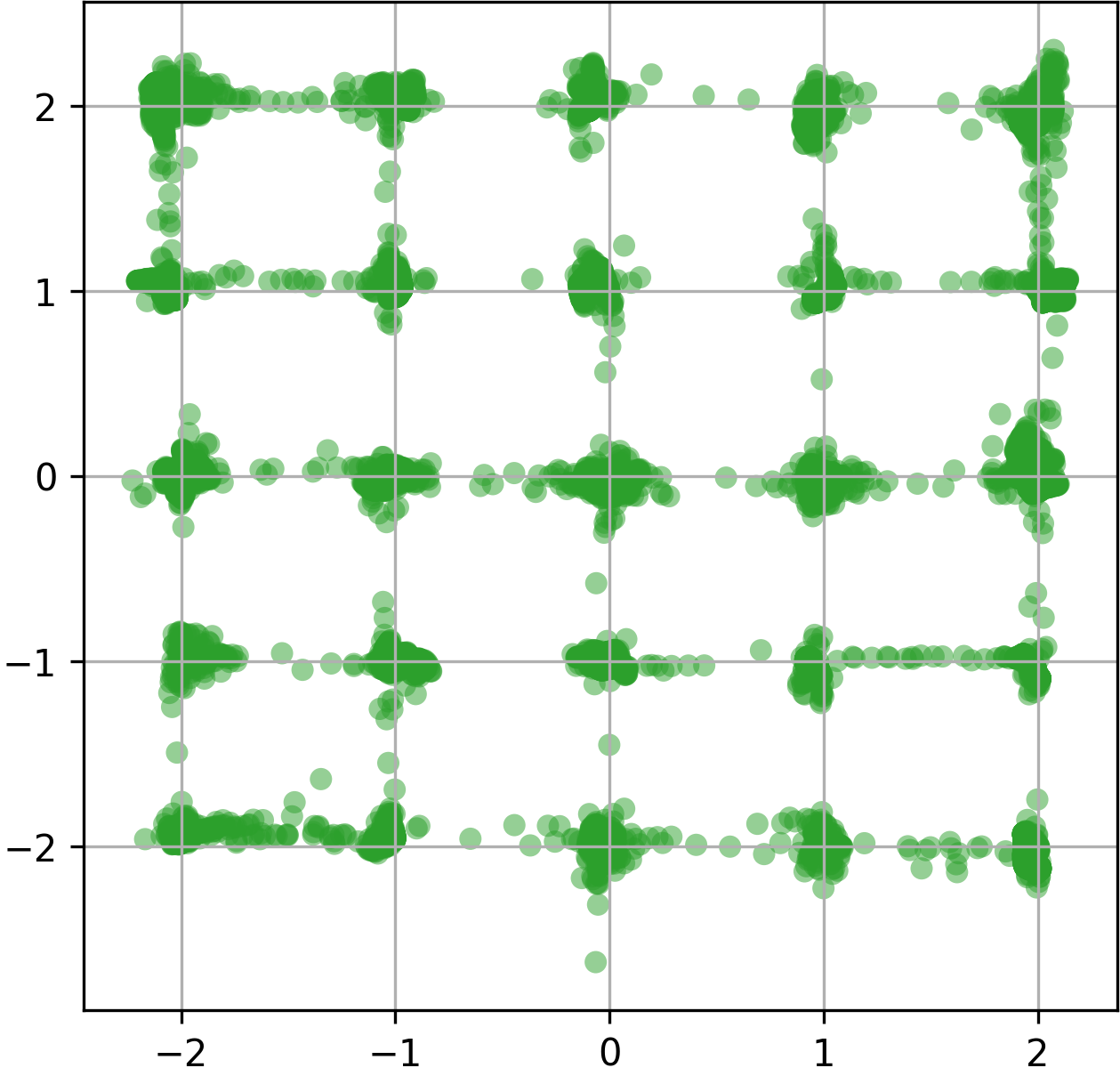}
    \label{fig:MH_samples}
    }
    \caption{Difference of distribution between training data and generators. The percentage of ``high-quality" samples for Figure~\ref{fig:direct_samples_5k}, Figure~\ref{fig:direct_samples_10k}, Figure~\ref{fig:direct_samples_50k} and Figure~\ref{fig:MH_samples} is 94.36\%, 94.31\%, 94.15\% and 95.64\%, respectively. The more we query, the more bad-quality samples we obtain, which affects the performance of model extraction. But if we reduce the number of queries, the performance of attack models still be poor due to insufficient training samples.}
    \label{fig:toy_example}
\end{figure*}

The reason why there exists a gap between the attack model and the target model in terms of {\it fidelity} is that the target GAN model is hard to reach global equilibrium and the discriminator is often better than the generator in practice~\cite{DRS2018discriminator}. As a result, real data distribution $p_r$ is not completely learned by the generator of the target model, which means that $p_g \neq p_r$.  Therefore, directly using the generator's distribution $p_g$ does not guarantee the high fidelity and it only minimizes the distribution discrepancy between the attack model and the target model. We explain this by a simple example on Figure~\ref{fig:toy_example}, which is popular in the GAN literature~\cite{DRS2018discriminator,MHGAN2019,DRE2020subsampling}.
 
Figure~\ref{fig:real_samples} presents real samples drawn from a mixture of 25 two-dimensional Gaussian distributions (each with standard deviation~$\sigma$ of 0.05). Figure~\ref{fig:direct_samples_5k} - Figure~\ref{fig:direct_samples_50k} show samples which are generated by a target GAN with different queries. We define a generated sample as ``high-quality" if its Euclidean distance to its corresponding mixture component is within four standard deviations (4$\sigma$ = 0.2)~\cite{DRS2018discriminator}. The architecture and setup information of the target GAN is shown in Appendix~\ref{ssec:Implementation details}. Overall, we can observe that target GAN's distribution is not completely the same as the training set's distribution, which means that directly extracting a model from the generator of the target GAN makes its distribution similar to the target model's distribution rather than its training dataset's distribution.

In order to achieve fidelity extraction, we believe that adding additional background information is necessary. Specifically, we consider two scenarios where adversaries can query the target model and also have limited auxiliary knowledge of:
\begin{enumerate}
	\item[(1)] partial real samples from training dataset; or
	\item[(2)] partial real samples and the discriminator of the target model.
\end{enumerate}
Partial real data can provide some information about distribution of the real data set, which can correct the fidelity extraction and make it closer to the real distribution. The discriminator from the target model can reveal the distribution information of the training data~\cite{DRS2018discriminator}. Thus, using the information provided by discriminator, we can subsample the generated data to make the obtained data closer to the real dataset's distribution, which benefits fidelity extraction. In Figure~\ref{fig:MH_samples}, we can observe that samples which generated by the MH subsampling algorithm~\cite{MHGAN2019} are much closer to the true distribution than those obtained by direct sampling. Specifically, it improves the percentage of ``high-quality" samples from 94.15\% to 95.64\%.

In fidelity extraction, we choose the same attack models and target models as accuracy extraction~(see Section~\ref{ssec:Target models and attack models}). In the following section, we call the first scenario of fidelity extraction as partial black-box fidelity extraction and the second scenario as white-box fidelity extraction.

\subsection{Methodology}
\label{ssec:Fidelity Methodology}
For partial black-box fidelity extraction, we first query the target model to obtain generated samples. And then, we train an attack model on these generated samples and continue training after adding partial real data. In this scenario, we assume that adversaries query the target model 50k times to obtain 50k generated samples. 

For white-box fidelity extraction, we first leverage the discriminator of the target model to subsample the generated samples. As a result, these refined samples are much closer to the true distribution. In this work, we use Metropolis-Hastings~(MH) subsampling algorithm~\cite{MHGAN2019} to subsample the generated data. See Algorithm~\ref{alg:mh subsampling} in Appendix for details. MH subsampling algorithm utilizes the discriminator through Metropolis-Hastings algorithm~\cite{tierney1994markov} to refine samples which are generated by the generator. The discriminator generally needs to be calibrated by partial real samples from training set of the target GAN model, considering that some discriminators of GANs output a score rather than a probability. In our experiments, all discriminators are calibrated through logistic regression. Then we train the attack model on those refined samples. After the training process of the attack model is stable, we add partial real data to further train the attack model. 

In this scenario, although the number of queries will increase due to subsampling samples, we assume that adversaries eventually obtain 50k refined samples. Partial real samples used to calibrate the discriminator are fixed to 10\% knowledge of training data. In addition, these partial real samples will be added into training process of the attack models.     

It is worth noting that we cannot directly choose the lowest {\it fidelity} value in real attack scenarios due to unavailability of training dataset from target models. Therefore, the {\it fidelity} value reported in this paper is chosen when its corresponding $\it accuracy$ value is the lowest in the training process.

\subsection{Results}
\subsubsection{\textbf{Partial black-box fidelity extraction.}} 
\begin{figure}[!t]
\centering
  \subfigure[Partial black-box fidelity extraction on CelebA]{
	\includegraphics[width=0.90\columnwidth]{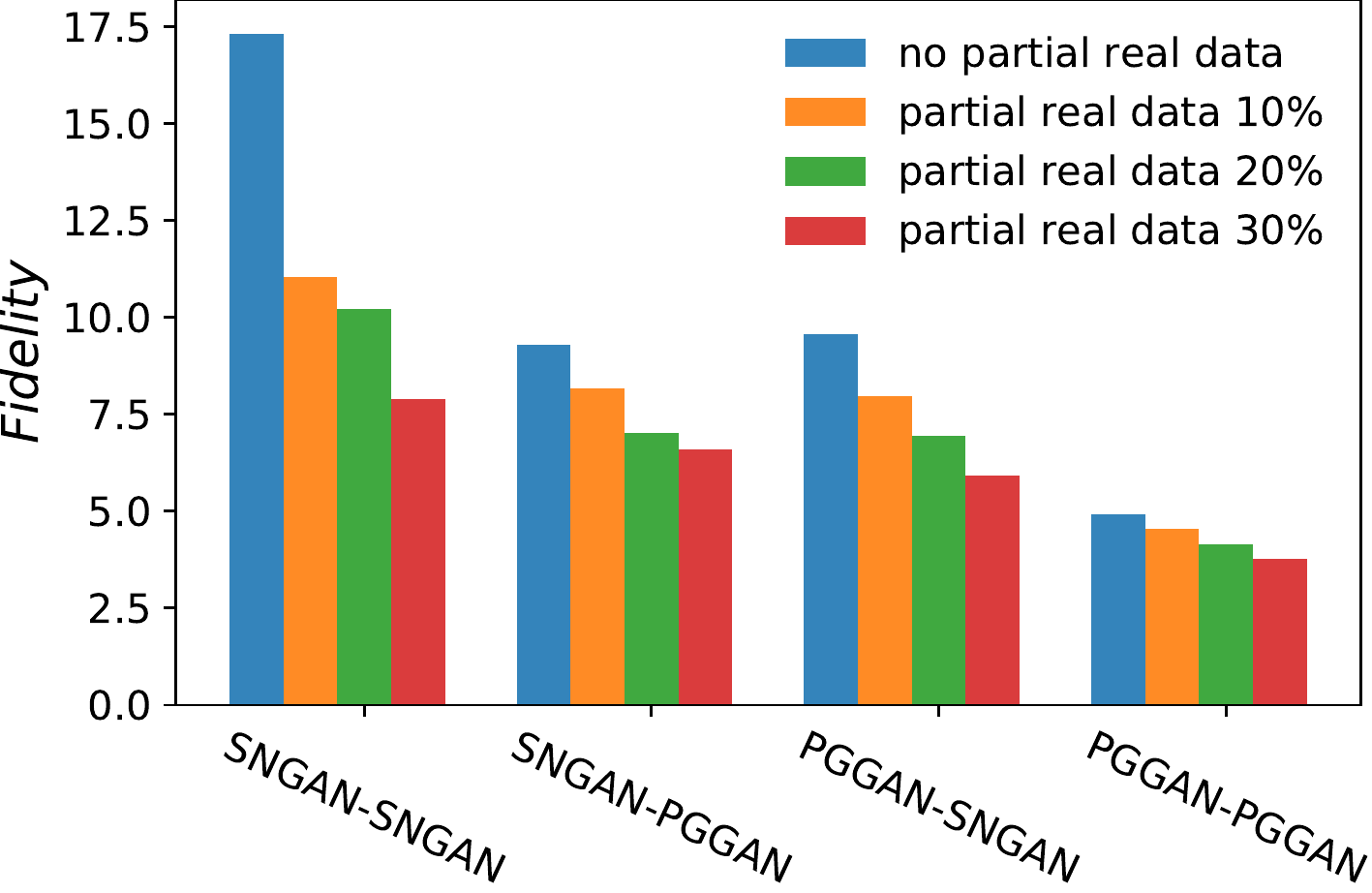}
	\label{fig:partial_black_box_celeba}
    }
  \subfigure[Partial black-box fidelity extraction on LSUN-Church]{
	\includegraphics[width=0.90\columnwidth]{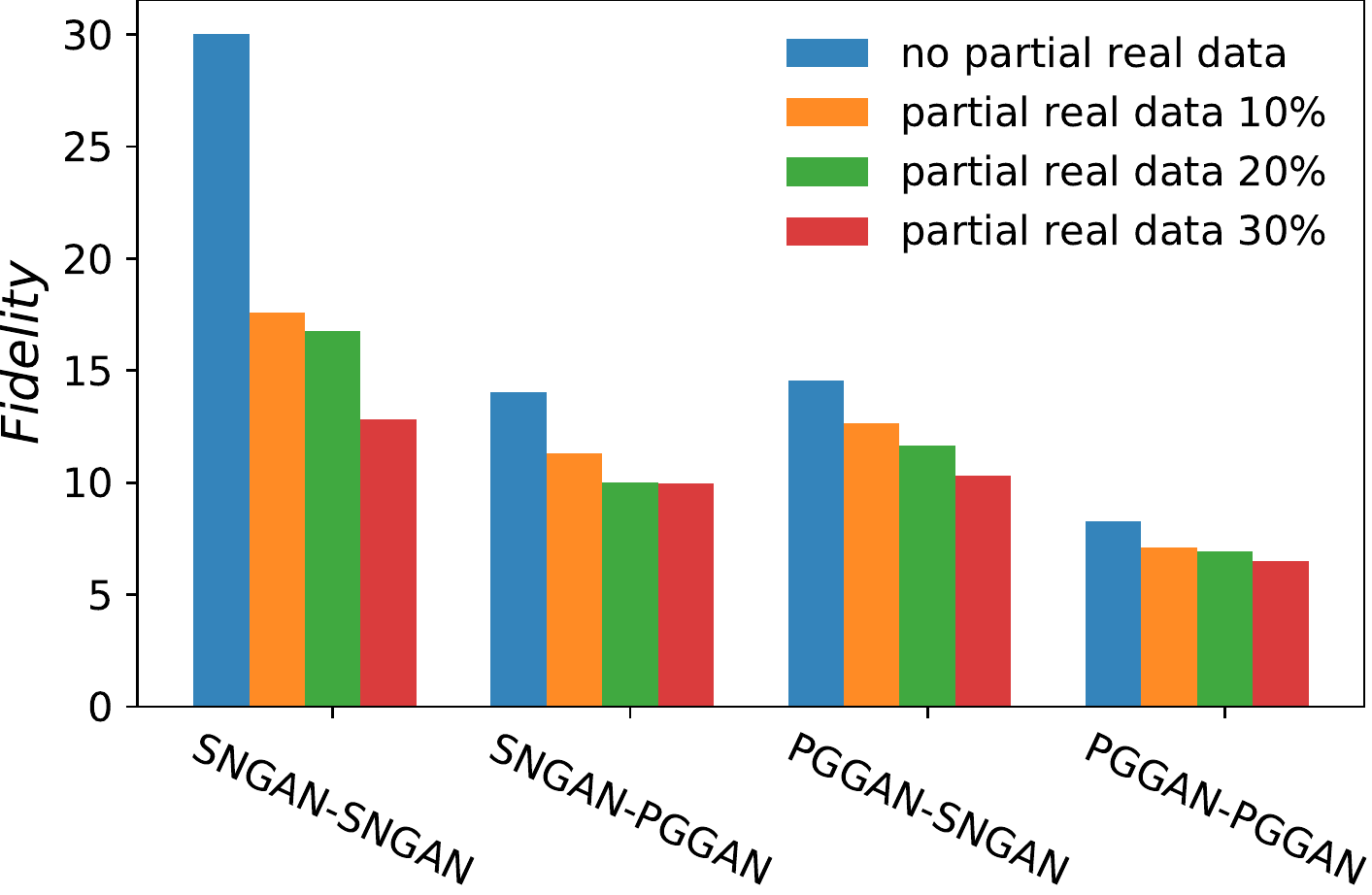}
	\label{fig:partial_black_box_church}
    }
    \caption{Partial black-box fidelity extraction.}
    \label{fig:partial_black_box}
\end{figure}

Figure~\ref{fig:partial_black_box_celeba} plots the results of the partial black-box fidelity extraction attack against target GAN models trained on CelebA dataset. We observe that directly adding the real data to a training dataset is an effective approach to improve {\it fidelity} of model extraction attacks for four different attack-target cases. For instance, compared to the SNGAN-SNGAN trained on CelebA with no partial real data, {\it fidelity} values decrease from 17.32 to 11.04 when adversaries have only 10\% knowledge of the training data from the target model. If adversaries gain 30\% knowledge of real data, {\it fidelity} further decreases to 7.89. That reminds model providers that training dataset should be kept secretly. In Figure~\ref{fig:partial_black_box_church}, we report the results of the partial black-box fidelity extraction attack against target GAN models trained on LSUN-Church dataset. Similar to Figure~\ref{fig:partial_black_box_celeba}, {\it fidelity} of attack models decreases with an increase with partial real data. Note that the significant decrease in {\it fidelity} of attack models can be observed between no partial real data and partial real data 10\% for all cases on both datasets, 
while the decrease in {\it fidelity} of attack models among partial real data 10\%, 20\%, and 30\% gradually becomes flat.

It is worth noting that increasing the number of queries can also improve {\it fidelity}, as shown in Figure~\ref{fig:fidelity_different_sizes}. However, we observe that it will gradually saturate while adding partial real data can break the limit. Table~\ref{tab:compare} compares the {\it fidelity} between two methods: increasing the number of queries and adding partial real data. We fix PGGAN trained on CelebA dataset as our target model and its FID is 3.40 which is also the ideal {\it fidelity} for attack models. For CelebA dataset, 10\% of the training set corresponds to 20,259 out of 202,599 images. Therefore, the size of training set of black-box accuracy extraction-70k  is roughly equal to that of partial black-box fidelity extraction 10\%. We can observe that adding equal amount of real data is more beneficial to {\it fidelity}, compared to increasing the number of queries. Furthermore, directly adding partial real data can break the limit of accuracy extraction. For instance, for the attack model PGGAN, obtaining 30\% of knowledge of real dataset can achieve 3.78 of {\it fidelity}, which is much closer to the ideal fidelity of 3.40, whereas increasing the number of queries to 90k still remains 4.44 of {\it fidelity}. Similar results can also be seen in attack model SNGAN shown in Table~\ref{tab:compare}, although {\it fidelity} of the attack model SNGAN is inferior to that of the attack model PGGAN.   

\begin{table}[!t]	
	\caption{Comparison between two methods: increasing the number of queries and adding partial real data.}
	\label{tab:compare}
	\centering
	\renewcommand{\arraystretch}{1.3}
	\scalebox{0.85}{
	\begin{tabular}{llr}
		\toprule
		Attack model & Method & {\it Fidelity}  \\
		&  &FID ($\Tilde{p}_g$, $p_r$)  \\
		\midrule
		\multirow{6}{*}{PGGAN}&Black-box accuracy extraction-50k& 4.93\\
		& Black-box accuracy extraction-70k& 4.65\\
		& Black-box accuracy extraction-90k& 4.44\\
		&Partial black-box fidelity extraction 10\%& 4.54\\
		&Partial black-box fidelity extraction 20\%& 4.15\\
		&Partial black-box fidelity extraction 30\%& 3.78\\
		\cline{1-3}
		\multirow{6}{*}{SNGAN}&Black-box accuracy extraction-50k& 9.29\\
		& Black-box accuracy extraction-70k& 8.38\\
		& Black-box accuracy extraction-90k& 7.38\\
		&Partial black-box fidelity extraction 10\%& 8.16\\
		&Partial black-box fidelity extraction 20\%& 7.02\\
		&Partial black-box fidelity extraction 30\%& 6.59\\
		\bottomrule
	\end{tabular}}
\end{table}

\subsubsection{\textbf{White-box fidelity extraction.}} As discussed in Section~\ref{ssec:Fidelity Methodology}, we also consider white-box fidelity extraction where adversaries obtain both partial real data and the discriminator of target models. Leveraging the discriminator of target models to subsample needs some real data from training dataset. These real data, on the one hand, is utilize to calibrate the discriminator which is illustrated in Algorithm~\ref{alg:mh subsampling}, on the other hand, it can be added the training process of the attack models to further improve the {\it fidelity} of attack models. We refer the former where only refined samples are used to train the attack model to MH fidelity extraction which is also considered as an indicator to show how well these refined samples are beneficial to {\it fidelity}. We refer the latter where both refined samples and partial real data are used to train the attack model as white-box fidelity extraction.     

\begin{figure}[!t]
\centering
  \subfigure[{\it Fidelity} on CelebA]{
	\includegraphics[width=0.90\columnwidth]{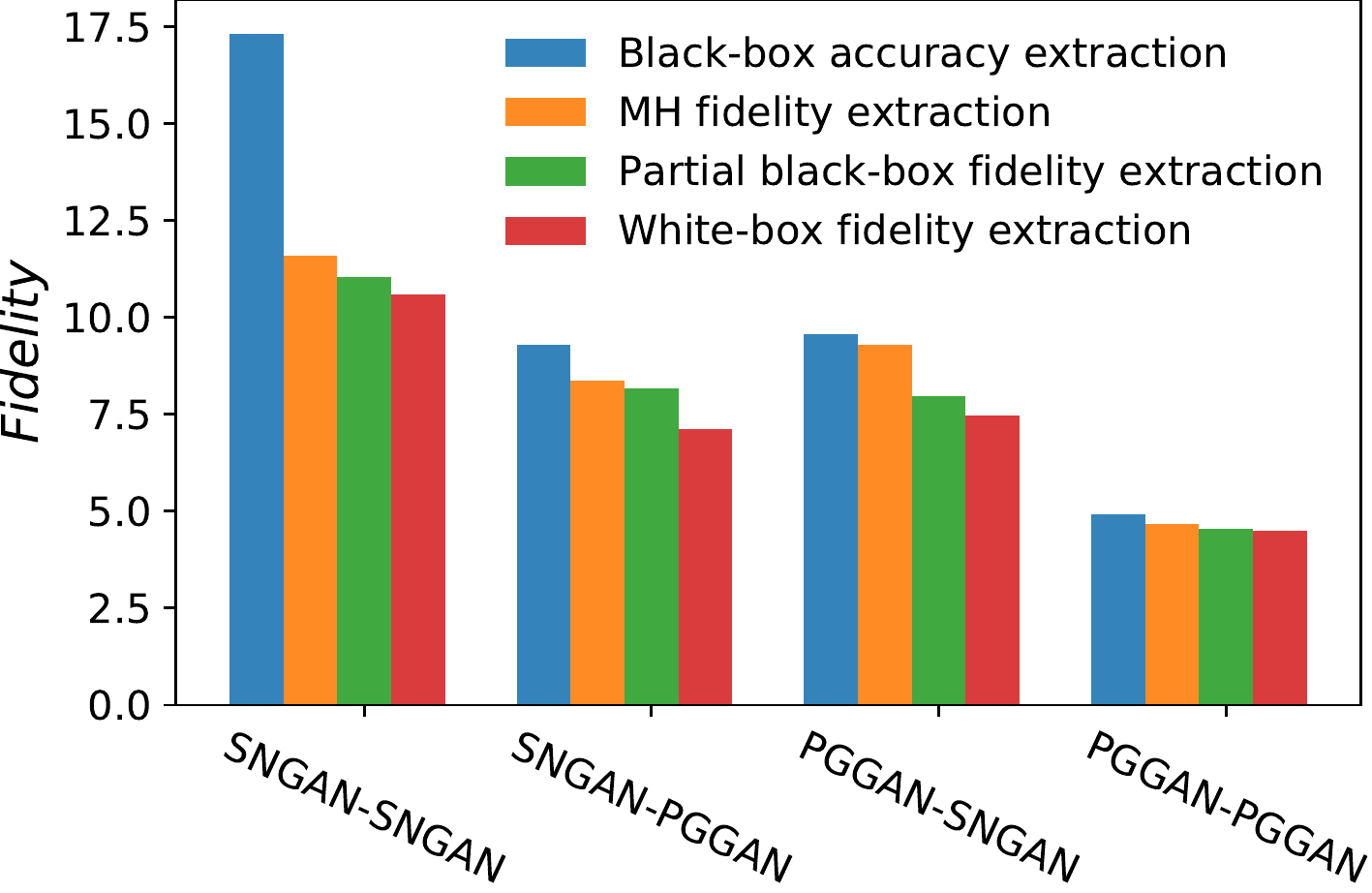}
	\label{fig:fidelity_celeba}
    }
  \subfigure[{\it Fidelity} on LSUN-Church]{
	\includegraphics[width=0.90\columnwidth]{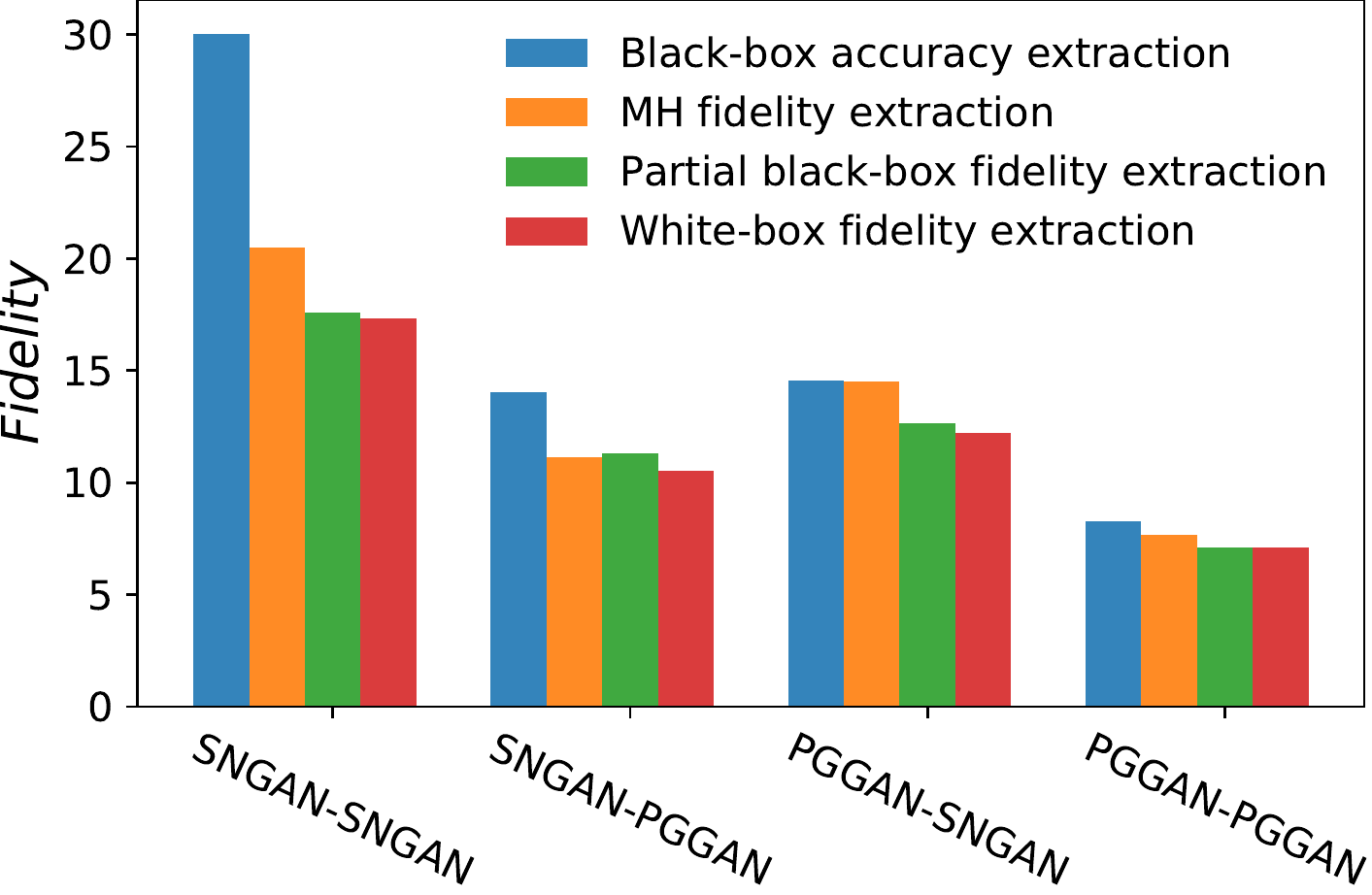}
	\label{fig:fidelity_church}
    }
    \caption{Comparison on {\it fidelity}.}
    \label{fig:comparison_all}
\end{figure}

Figure~\ref{fig:comparison_all} plots not only the results of MH fidelity extraction and white-box fidelity extraction on both CelebA and LSUN-Church datasets, but also black-box accuracy extraction and partial black-box fidelity extraction for comparison. We can observe that MH subsampling is an effective approach to improve {\it fidelity} of attack models. For example, when target model is SNGAN, MH fidelity extraction can significantly improve attack model's fidelity on both datasets because MH subsampling algorithm selects high-quality samples from generated samples of the target model SNGAN. For partial black-box fidelity extraction and white-box fidelity extraction which both leverage the partial real data in the training process, white-box fidelity extraction generally can achieve higher fidelity than partial black-box fidelity extraction. We also analyze distribution differences for fidelity extraction, which is shown in Figure~\ref{fig:dissect_fidelity} in Appendix.

%% file: 7_case_study.tex
\section{Case study: model extraction based transfer learning}  
\label{sec:casestudy1}
In this section, we present one case study where the extracted model serves as a pretrained model and adversaries transfer knowledge of the extracted model to new domains by means of fine-tuning to broaden the scope of applications based on extracted models. We start with methods of transfer learning on GAN and demonstrate how adversaries can benefit from model extraction, in addition to directly leveraging the extracted model to generate images.  

We consider the state-of-the-art GAN model StyleGAN~\cite{StyleGAN12019style} that was trained on more than 3 million bedroom images as the target model. StyleGAN produces high-quality images at a resolution of $256 \times 256$, with 2.65 FID on LSUN-Bedroom dataset~\cite{yu2015lsun}. We suppose adversaries only query the target model StyleGAN and have no any other background knowledge, which is also called black-box accuracy extraction in our paper. Although an adversary can obtain an extracted model, the model only generates images which are similar to the target model. In this case, the extracted model can only generate bedroom images due to target model trained on LSUN-Bedroom dataset. Therefore, the adversary's goal is to use the PGGAN as the attack model to extract the target model StyleGAN and leverage transfer learning to obtain a more powerful GAN which generates images that the adversary wishes. \textit{The attack is successful if the performance of models training by transfer learning based on the extracted GAN outperforms models training from scratch.}

Transferring knowledge of models which steal the state-of-the-art models to new domains where adversaries wish the GAN model can generate other types of images can bring at least two benefits: 1) if adversaries have too few images for training, they can easily obtain a better GAN model on limited dataset through transfer learning; 2) even if adversaries have sufficient training data, they can still obtain a better GAN model through transfer learning, compared with a GAN model training from scratch. Therefore, we consider two variants of this attack: one where the adversary owns a small target dataset (i.e., about 50k images in our work) and the other one where the adversary has enough images (i.e., about 168k images in our work).  

More specifically, after querying the target model StyleGAN and obtaining 50k generated images, adversaries train their attack model PGGAN on the obtained data, as illustrated in Section~\ref{ssec:Accuracy Methodology}. {\it Accuracy} of the attack model PGGAN is 4.12 and its {\it fidelity} is 6.97. Then, we use the extracted model's weights as an initialization to train a model on adversary's own dataset which is also called target dataset in the section. We conduct the following two experiments:

\begin{enumerate}
	\item We first randomly select 50k images from LSUN-Kitchen dataset as a limited dataset. Then, we train the model on these selected data by transfer learning and from scratch, respectively.    
	\item We train a model on the LSUN-Classroom dataset including about 168k images by transfer learning and from scratch, respectively.
\end{enumerate}

\begin{figure}[!t]
\centering
  \subfigure[FID on LSUN-Kitchen]{
	\includegraphics[width=0.450\columnwidth]{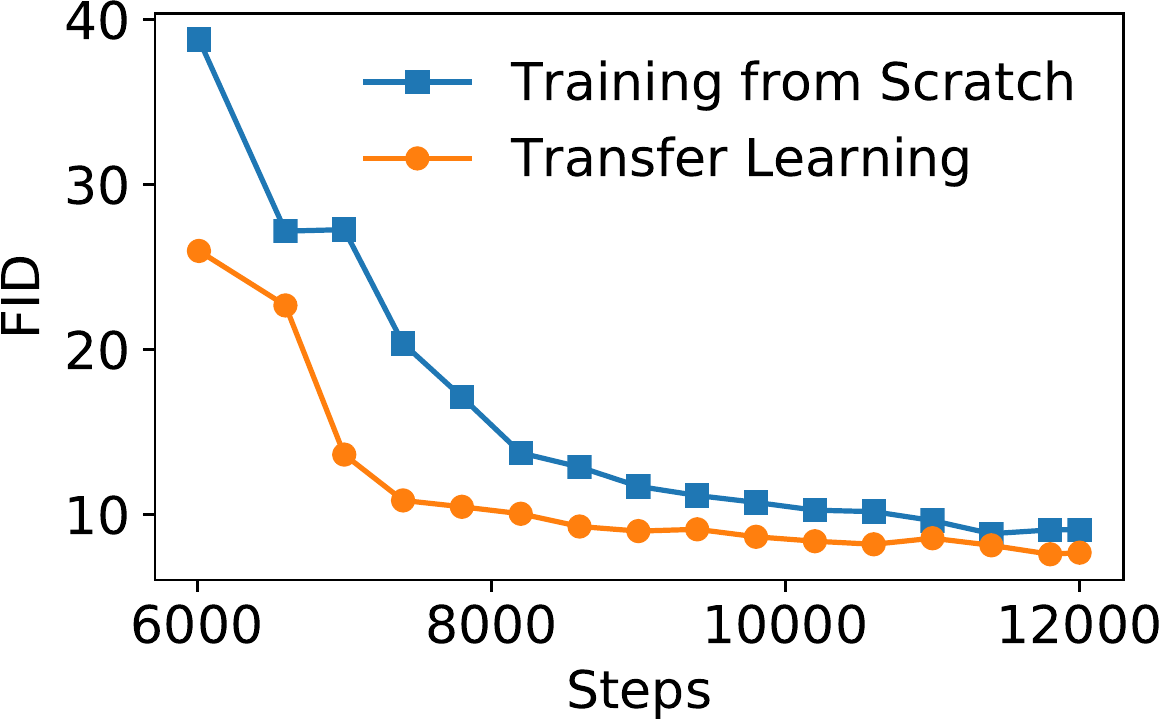}
	\label{fig:fid_kitchen}
    }
  \subfigure[FID on LSUN-Classroom]{
	\includegraphics[width=0.450\columnwidth]{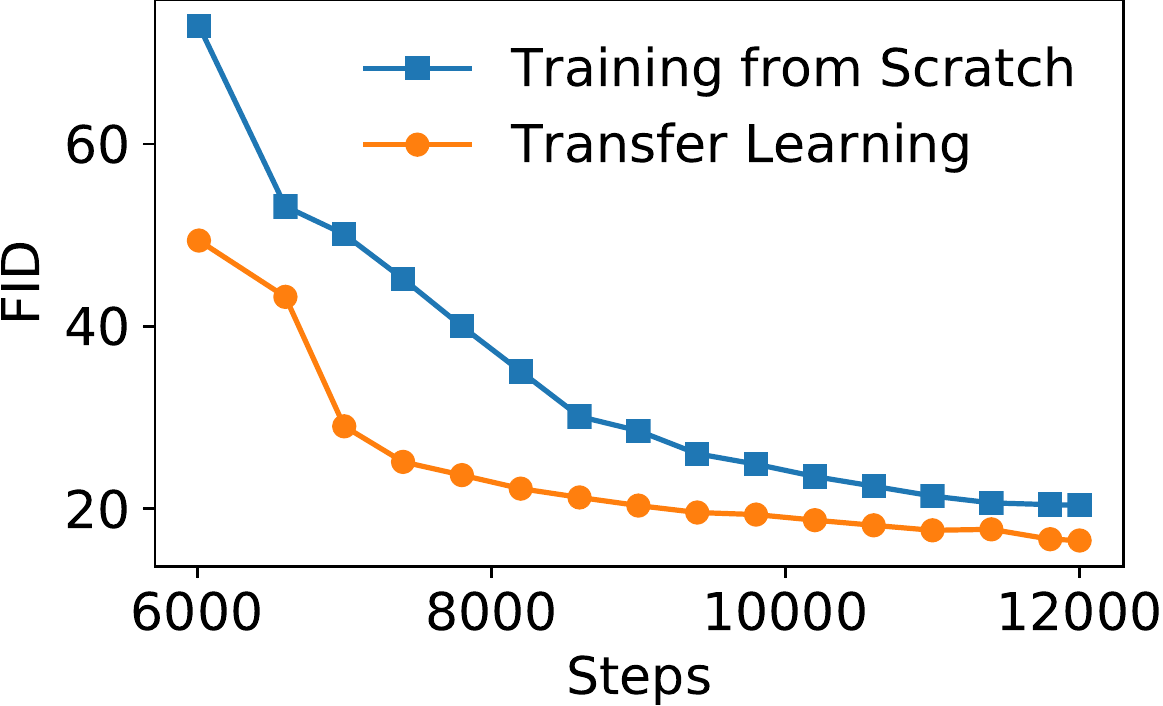}
	\label{fig:fid_classroom}
    }
    \caption{Comparison between transfer learning and scratch on LSUN-Kitchen and LSUN-Classroom dataset.}
    \label{fig:transfer_learning}
\end{figure}

\smallskip\noindent
\textbf{Results.} Table~\ref{tab:transfer_learning} shows the performance of models trained by transfer learning and training from scratch. We can observe that the performance of training by transfer learning is always better than that of training from scratch on both large and small target dataset. To be specific, on the limited LSUN-Kitchen dataset which contains 50k images, the FID of model trained by transfer learning decreases from 8.83 to 7.59, compared with the model trained from scratch. It indicates that the extracted model is useful for models trained on other types of images. On the large LSUN-Classroom dataset which contains more than 168k classroom images, the performance of model significantly improves from model training from scratch with 20.34 FID\footnote{This value is not equal to 20.36~\cite{PGGAN2018progressive} due to randomness.} to training by transfer learning with 16.47 FID. This is also the best performance for PGGAN on LSUN-Classroom dataset, in contrast with 20.36 FID reported by Karras et al.~\cite{PGGAN2018progressive}. We also plot the process of training for both settings on the two datasets, which is shown in Figure~\ref{fig:transfer_learning}. We can obviously and consistently observe that training by transfer learning is always better than training from scratch during the training process, which indicates that the extracted model PGGAN which duplicates the state-of-the-art StyleGAN on LSUN-Bedroom dataset can play a significant role in other applications rather than only on generating bedroom images. That reminds us that model extraction on GANs severely violates intellectual property of the model owners.

\begin{table}[!t]	
	\caption{Comparison between transfer learning and training from scratch. The target model is StyleGAN trained on LSUN-Bedroom dataset, and the attack model is PGGAN.}
	\label{tab:transfer_learning}
	\centering
	\renewcommand{\arraystretch}{1.1}
	\scalebox{1}{
	\begin{tabular}{llr}
		\toprule
		Target dataset & Methods & FID \\
		\midrule
		LSUN-Kitchen & Transfer Learning & 7.59\\
		LSUN-Kitchen & Training from Scratch & 8.83\\
		LSUN-Classroom & Transfer Learning & 16.47\\
		LSUN-Classroom & Training from Scratch & 20.34\\
		\bottomrule
	\end{tabular}}
\end{table}

%% file: 8_defense.tex
\section{Defenses}
\label{sec:defenses}
Model extraction attacks on GANs leverage generated samples from a target GAN model to retrain a substitutional GAN which has similar functions to the target GAN. In this section, we introduce defense techniques to mitigate model extraction attacks against GANs.

According to adversary's goals as defined in Section~\ref{ssec:adversarygoals}, we discuss defense measures from two aspects: {\it accuracy} and {\it fidelity}.
In terms of {\it accuracy} of model extraction, it is difficult for model owners to defend except for limiting the number of queries. This is because adversaries can always design an attack model to learn the distribution based on their obtained samples. The more generated samples adversaries obtain, the more effective they achieve. 

In terms of {\it fidelity} of model extraction, its effectiveness is mainly because adversaries are able to obtain samples generated by latent codes draw from a prior distribution of the target model, and these samples generated through the prior distribution are representative for real data distribution~\cite{agustsson2018optimal}. However, if adversaries obtain some generated samples which are only representative for partial real data distribution or a distorted distribution, {\it fidelity} of attack models becomes poor. Based on this, we propose two types of perturbation-based defense mechanisms: input perturbation-base and output perturbation-based approaches. In the rest of this section, we focus on defense approaches which are designed to mitigate {\it fidelity} of attack models.

\begin{figure}[!t]
	\centering
	\includegraphics[width=1\linewidth]{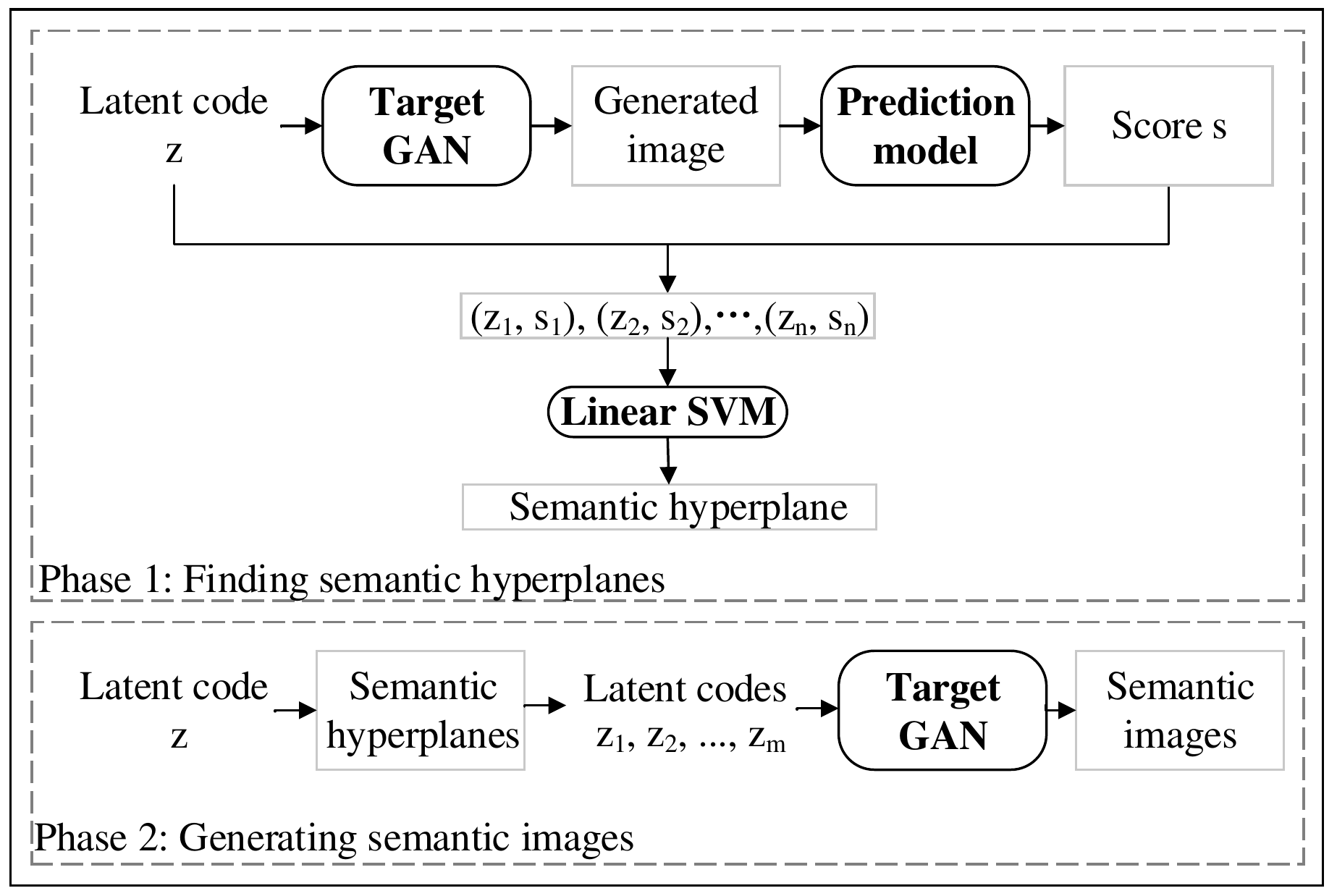}
	\caption{Semantic interpolation defense. }
	\label{fig:Semantic defense}
\end{figure}

\subsection{Methodology}
\label{ssec:defense Methodology}

\subsubsection{Input perturbation-base defenses.} For this type of defenses, we propose two approaches based on perturbing latent codes: linear interpolation defense and semantic interpolation defense.

\smallskip\noindent
\textbf{Linear interpolation defense.} For $n$ latent codes queried from users, model providers randomly select two queried points and interpolate $k$ points between the two points. This process is repeated for $\lceil n/k \rceil $ times to get $n$ modified latent codes. These modified latent codes are used to query the target model. In our experiments, we interpolate 9 points. See Figure~\ref{fig:linear9} in Appendix for visualization.

\smallskip\noindent
\textbf{Semantic interpolation defense.} Unlike linear interpolation defense where target models return a batch of random images, semantic interpolation defense enables users to obtain a batch of semantic images. Semantic information can be any information that humans can perceive, which is usually defined by model providers. For instance, for a human face image, semantic information includes gender, age and hair style. In this paper, we adopt a semantic interpolation algorithm proposed by Shen et al.~\cite{shen2020interpreting}. 

Specifically, the process of semantic interpolation defense is shown in Figure~\ref{fig:Semantic defense}. Semantic interpolation defense consists of two phases: finding semantic hyperplanes and generating semantic images. In the first phase, we first train a prediction model for each semantic information. Then the trained prediction model is used to predict semantic score $s$ for each image generated through latent code $z$. As a result, we get latent code-score pairs and label the highest $k$ scores as positive and the lowest $k$ scores as negative. Finally, we train a linear support vector machine~(SVM) on dataset where latent codes as training data and scores as labels. A trained linear SVM contains a hyperplane which separates one semantic information. In the second phase, we can obtain a semantic image for each semantic hyperplane through interpolation. A latent code interpolates points along the normal vector of the hyperplane and corresponding semantic images can be obtained. In our experiments, we totally explore 12 semantic information on CelebA dataset. See Figure~\ref{fig:sematic} in Appendix for visualization.

In our experiments, we train each prediction model for each semantic information, and prediction model is built on the basis of ResNet-50 network~\cite{he2016deep} trained on ImageNet dataset~\cite{russakovsky2015imagenet}. 

\subsubsection{Output perturbation-base defenses.}
Instead of perturbing latent codes, this type of defenses directly perturbs the generated samples. Specifically, we propose four approaches: random noise, adversarial noise, filtering and compression. See Figure~\ref{fig:output_defense_images} in Appendix for visualization.

\smallskip\noindent
\textbf{Random noise.} Adding random noises on generated samples is a straightforward method. In our experiments, we use Gaussian-distributed additive noises~(mean = 0, variance = 0.001).   

\smallskip\noindent
\textbf{Adversarial noise.} We generate adversarial examples through mounting targeted attacks where all images are misclassified into a particular class by the classifier ResNet-50 trained on ImageNet dataset. 
In our experiments, all face images are misclassified into the class --- goldfish and the C\&W algorithm~\cite{carlini2017towards} based on $L_2$ distance are used. 

\smallskip\noindent
\textbf{Filtering.} The Gaussian filter is used to process generated samples. In our experiments, we use Gaussian filter~(sigma = 0.4) provided by the skimgae package~\cite{scikit-image}.

\smallskip\noindent
\textbf{Compression.} The JPEG compression algorithm is used to process generated samples. In our experiments, we use the JPEG compression~(quality = 85) provided by the simplejpeg package~\cite{simplejpeg}.

\begin{figure}[!t]
\centering
    \includegraphics[width=1\columnwidth]{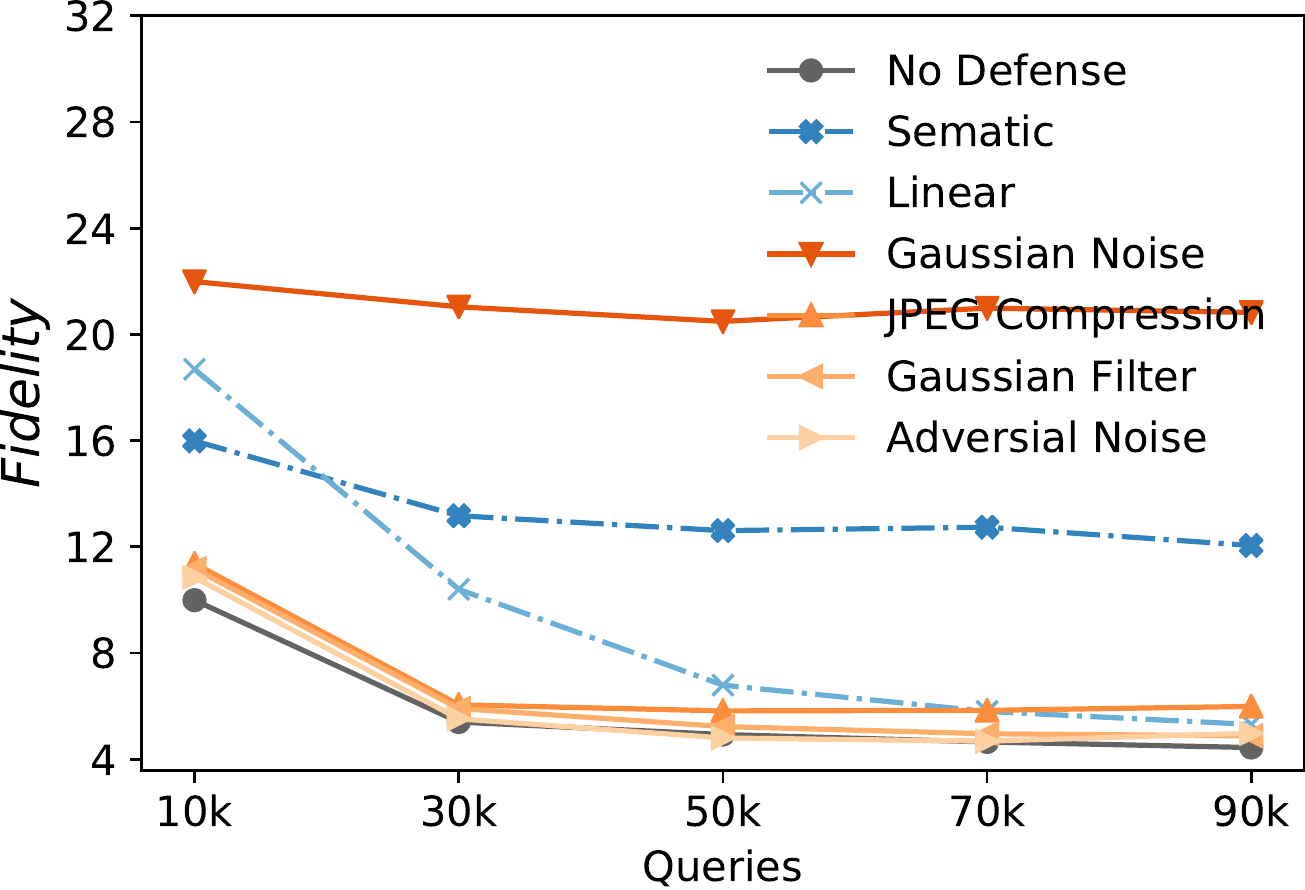}
    \caption{The performance of attack model PGGAN for defenses on black-box accuracy extraction.}
    \label{fig:defense_pggan}
\end{figure}

\subsection{Results}
\label{ssec:defense results}

In this experiment, we choose PGGAN trained on CelebA dataset as the target model to evaluate our defense techniques, considering its excellent performance among our target models. 
We only show the effectiveness of defense techniques on black-box accuracy extraction, considering its more practical assumption: adversaries obtain samples by model providers or queries.

\subsubsection{\textbf{Defense on black-box accuracy extraction.}} Figure~\ref{fig:defense_pggan} plots results of attack model PGGAN on defenses. We observe that attack performance is weakened when the target model PGGAN uses these defense approaches, compared to the target model without any defense. 
Gaussian noise and semantic interpolation show stable performance while other defense techniques' performance is weakened with an increase in the number of queries. Figure~\ref{fig:defense_sngan} in Appendix also shows similar defense performance for the attack model SNGAN.
 
\subsubsection{\textbf{Discussion.}} 
The reason why input perturbation-based defenses can work is at least explained from two aspects: increasing the similarity of generated samples and a distribution mismatch between latent codes produced by interpolation and drawn from prior distribution. For the former, we can see that interpolation operations increase the similarity of images from Figure~\ref{fig:input_defense_images}. For the latter, latent codes produced by interpolation operations are different from latent codes drawn from the prior distribution that the target model was trained on. This is because latent codes produced by linear operation do not obey the prior distribution of the target model, which also bring a benefit in disguising the true data distribution~\cite{agustsson2018optimal}.

Output perturbation-based defenses can work because they directly perturb these generated samples. Model providers need to trade-off image quality and the model's security through magnitudes of changes. Although Gaussian noise defense shows the best performance, it is possible for adversaries to remove noise.

%% file: 9_conclusion.tex
\section{Conclusion}
\label{sec:conclusion}
In this paper, we have systematically studied the problem of model extraction attacks on generative adversarial networks, and devised, implemented, and evaluated this attack from the perspective of accuracy extraction and fidelity extraction. For accuracy extraction, extensive experimental evaluations show that adversaries can achieve an acceptable performance with about 50k queries. For fidelity extraction, adversaries further improve the {\it fidelity} of attack models after obtaining additional background knowledge, such as partial real data from training set or the discriminator of the target model. We have also performed a case study where the attack model which steals a state-of-the-art target model can be transferred to new domains to broaden the scope of applications based on extracted models. Finally, we proposed two types of effective defense techniques based on perturbing latent codes or generated samples to mitigate model extraction attacks:
input and output perturbation-based defense. Semantic interpolation and Gaussian noise defenses show a stable performance.

Training with differential privacy techniques can be utilized to protect privacy of training data of a model~\cite{abadi2016deep}. However, training time and stability of the training process are big challenges for GANs. For further work, we plan to design new methods based on differential privacy techniques to mitigate {\it fidelity} of model extraction.
Additionally, studying watermarks to defend model extraction against GANs is also an interesting direction for future work.

%% file: appendix.tex
\section{Appendix}
\subsection{Implementation details}
\label{ssec:Implementation details}

We implement PGGAN\footnote{\url{https://github.com/tkarras/progressive_growing_of_gans}} and SNGAN\footnote{\url{https://github.com/christiancosgrove/pytorch-spectral-normalization-gan}} based on following codes indicated in the footnotes.
We choose the ResNet architecture for SNGAN and the architecture of PGGAN is the same as the official implementation.
We use hinge loss for SNGAN and WGAN-GP loss for PGGAN. For target GAN on synthetic data in Figure~\ref{fig:toy_example}, we use four fully connected layers with ReLU activation for both generator and discriminator and the prior is a 2-dimensional standard normal distribution. The training data is a mixture of 25 2-D Gaussian distributions (each with standard deviation of 0.05). We train it using standard loss function~\cite{goodfellow2014generative}. In Section~\ref{sec:casestudy1} about case study,  
we directly use the pretrained StyleGAN\footnote{\url{https://github.com/NVlabs/stylegan}} trained on LSUN-Bedroom dataset as our target model. We resize all images used in our paper to $64 \times 64$, except for the case study where images with a resolution of $256 \times 256$ are used. The dimension of latent space of SNGAN, PGGAN and StyleGAN is 256, 512 and 512, respectively, and their latent codes are all draw from standard Gaussian distribution. 

In Section~\ref{sec:defenses} about semantic interpolation defense, the semantic information is from attributes of CelebA dataset\footnote{\url{http://mmlab.ie.cuhk.edu.hk/projects/CelebA.html}}, which has labeled for each image. we only choose 12~(male, smiling, wearing lipstick, mouth slightly open, wavy hair, young, eyeglasses, wearing hat, black hair, receding hairline, bald, mustache) out of 40 facial attributes to learn semantic hyperplanes, because the number of images for each attribute varies largely and some attributes is hard to distinguish when they are applied in target GAN model. We train prediction model for each attribute based on ResNet-50 model pretrained on ImageNet\footnote{\url{https://download.pytorch.org/models/resnet50-19c8e357.pth}}.

\subsection{MH algorithm}
\label{ssec:MH algorithm}
Algorithm~\ref{alg:mh subsampling} shows the MH subsampling algorithm. Inputs of this algorithm are a target generator~(only used to query), a white-box discriminator which is used to subsample generated samples and partial real samples which are used to calibrate the discriminator~(Algorithm~\ref{alg:mh subsampling}, line 2). Outputs are refined samples whose distribution is much closer to distribution of real training data. 

\renewcommand{\algorithmicrequire}{ \textbf{Input:}}
\renewcommand{\algorithmicensure}{ \textbf{Output:}} 
\begin{algorithm}[tb]
   \caption{MH subsampling}
   \label{alg:mh subsampling}
\begin{algorithmic}[1]
	\REQUIRE target generator $G$, target discriminator $D$,\\~~~~~ partial real samples $X_r = \{x_{r1}, x_{r2}, $\dots$, x_{rm}\}$
	\ENSURE $N$ refined images 
	\STATE Sample $m$ fake images $X_g = \{x_{g1}, x_{g2}, $\dots$, x_{gm}\}$ from $G$
	\STATE Train a calibrated classifier: \\
	$C \leftarrow LogisticRegression(D(X_r), D(X_g))$  
	\STATE $images\leftarrow \emptyset$\
   	\WHILE {$|images|<N$}
   	\STATE $x \leftarrow$ a real image from $X_r$
   	\FOR{$i=1$ to $K$}
   	\STATE Sample $x'$ from $G$
   	\STATE Sample $u$ from Uniform$(0,1)$
   	\STATE Compute real image's density ratio: \\ $r(x) = \frac{C(D(x))}{1-C(D(x))}$
   	\STATE Compute fake image's density ratio: \\ $r(x') = \frac{C(D(x'))}{1-C(D(x'))}$
   	\STATE $p = min(1, \frac{r(x')}{r(x)})$
   	\IF{$u\leq p$}
   	\STATE $x \leftarrow x'$\ 
   	\ENDIF
   	\ENDFOR
   	\IF{$x$ is not a real images}
   	\STATE Append($x, images$)
   	\ENDIF
   	\ENDWHILE
\end{algorithmic}
\end{algorithm}

\subsection{More results for analyzing distribution differences}
\label{ssec:Distribution difference}
\subsubsection{\textbf{Understanding accuracy extraction for the target model SNGAN}} 
\begin{figure}[!ht]
\centering
	\includegraphics[width=1\columnwidth]{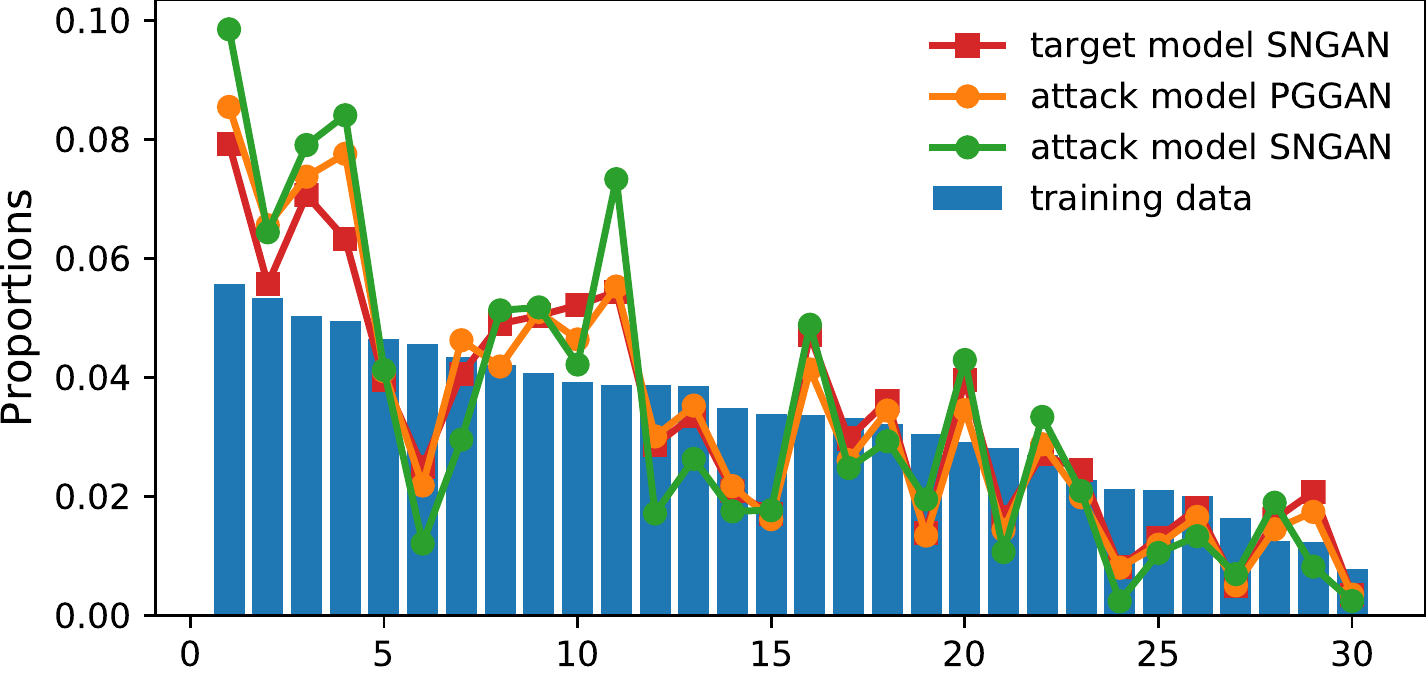}	
	\caption{Class distributions of the training data, the target model SNGAN, and attack models.}
	\label{fig:dissect_accuracy_sngan}
\end{figure}

Figure~\ref{fig:dissect_accuracy_sngan} shows distribution differences for the target model SNGAN trained on CelebA dataset. Table~\ref{tab:accuracy attack_js_sngan} summarizes these differences statistically.
\begin{table}[!t]
\centering
	\caption{JS distances between models. For the JS distance between training data and the target model, and the target model SNGAN is $16.36 \times 10^{-3}$.}	
	\label{tab:accuracy attack_js_sngan}
	\renewcommand{\arraystretch}{1.1}
	\scalebox{0.9}{
	\begin{tabular}{llrr}
		\toprule
		Target model & Attack model & $\it JS_{\it accuracy}$ ($\times 10^{-3}$)& $\it JS_{\it fidelity}$ ($\times 10^{-3}$)\\
		\midrule

		\multirow{2}{*}{SNGAN} &SNGAN &8.90 & 34.12\\	
		&PGGAN & 1.60 & 18.56\\

		\bottomrule
	\end{tabular}}
\end{table}

\subsubsection{\textbf{Understanding fidelity extraction on GANs in-depth.}}
\label{Understanding fidelity extraction}
\begin{figure}[!t]
\centering
	\subfigure[The target model PGGAN trained on CelebA.]{
    \includegraphics[width=1\columnwidth]{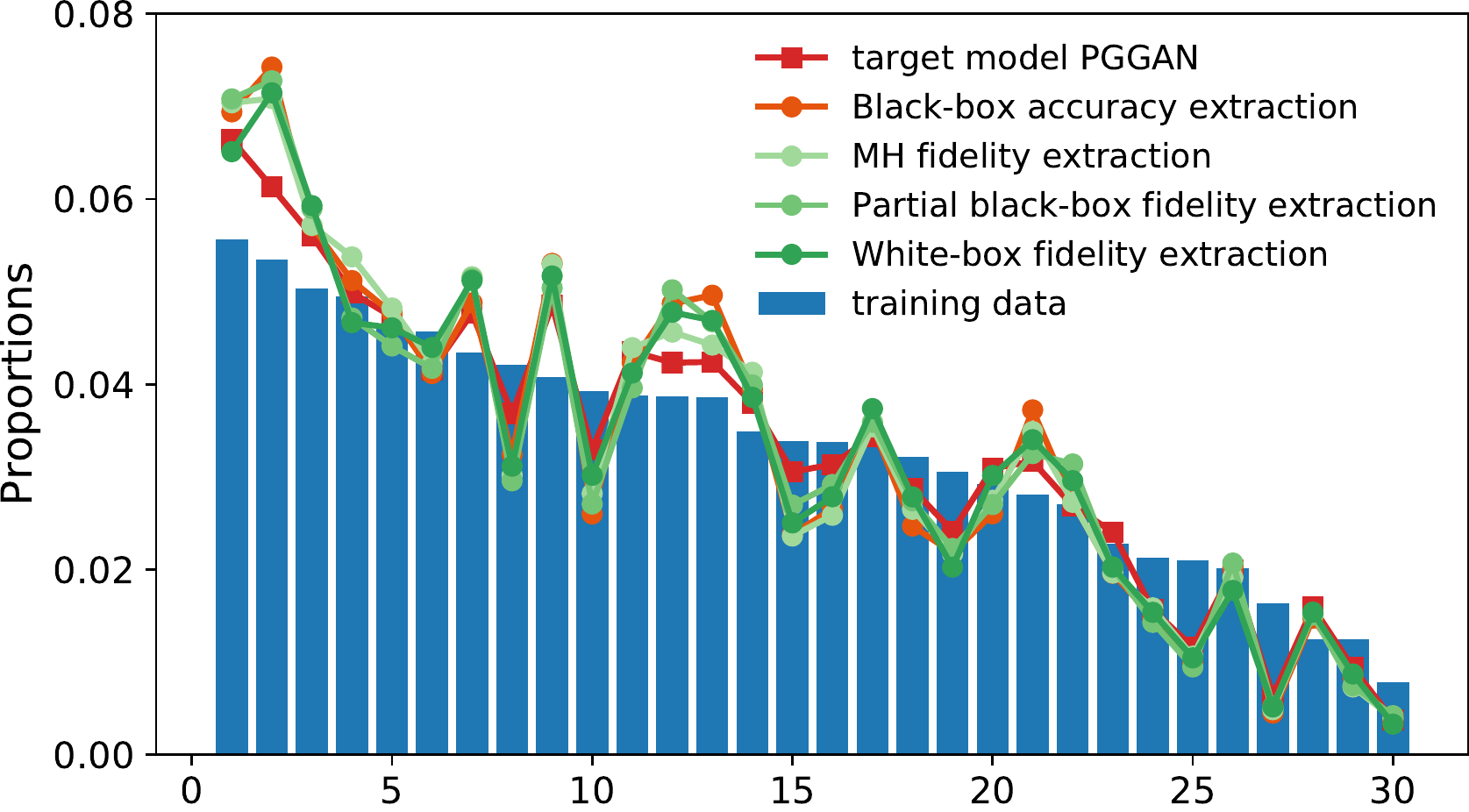}
    \label{fig:dissect_fidelity_celeba}
    }
    \subfigure[The target model PGGAN trained on LSUN-Church.]{
        \includegraphics[width=1\columnwidth]{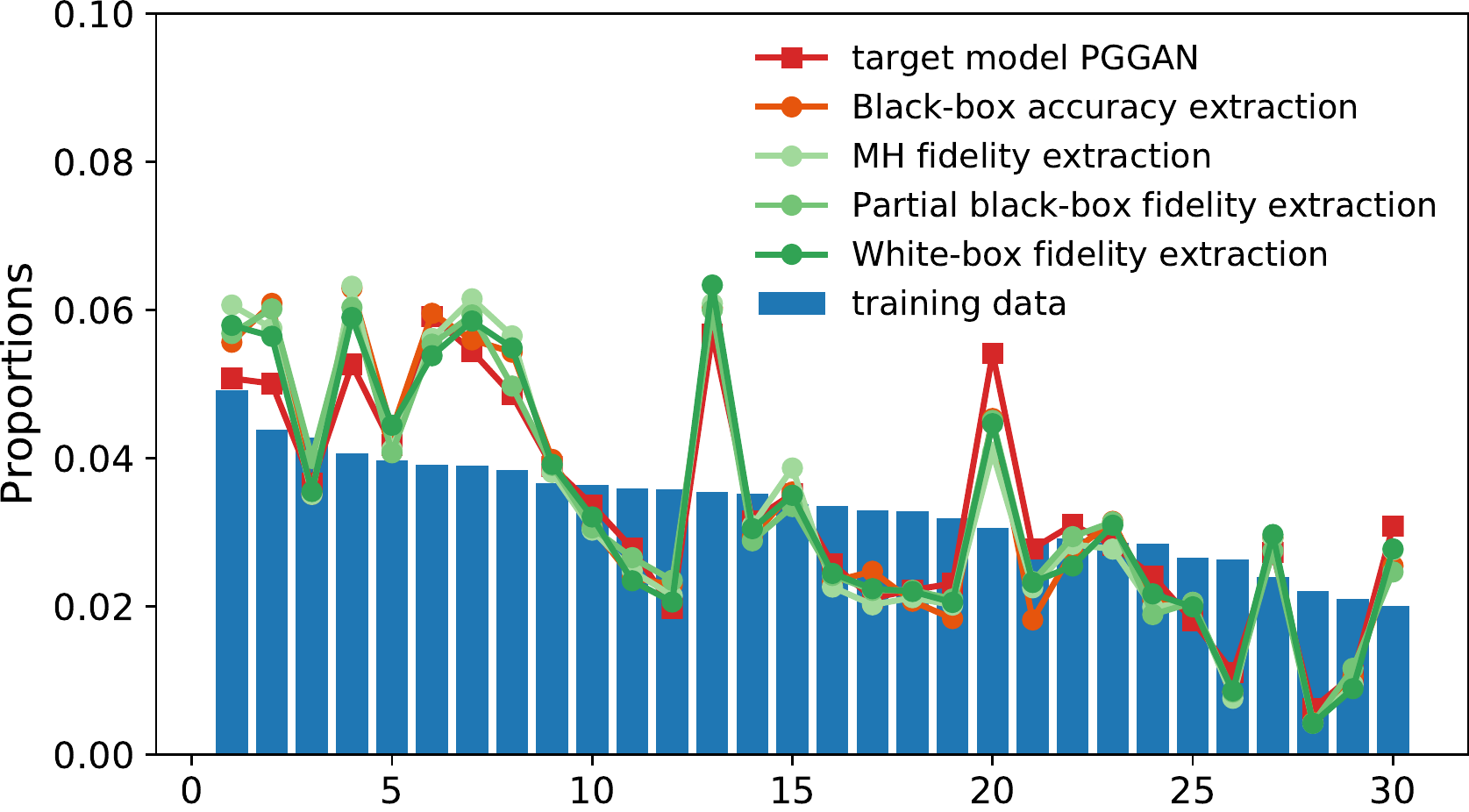}
        \label{fig:dissect_fidelity_church}
    }
    \caption{Distribution differences for fidelity extraction.}
    \label{fig:dissect_fidelity}
\end{figure}

Following the same procedure illustrated in Section~\ref{Understanding accuracy extraction}, we also dissect distribution differences for fidelity extraction.
Specifically, we choose the PGGAN-PGGAN case as an example~(see Figure~\ref{fig:comparison_all}) and the attack models is PGGAN. 
From the Figure~\ref{fig:dissect_fidelity}, we observe that for CelebA, white-box fidelity extraction which has minimal $\it fidelity$ values among these methods is more consistent with the distribution of the training data by lowering the highest proportions of classes. For LSUN-Church, similar results also can be observed.
Table~\ref{tab:fidelity attack_js} summarizes these differences statistically.

\begin{table}[!t]
\centering
	\caption{JS distances between models. For the JS distance between training data and the target model, the target model PGGAN on CelebA is $4.14 \times 10^{-3}$ and the target model PGGAN on LSUN-Church is $14.78 \times 10^{-3}$.}	
	\label{tab:fidelity attack_js}
	\renewcommand{\arraystretch}{1.3}
	\scalebox{0.7}{
	\begin{tabular}{llrr}
		\toprule
		Dataset & Methods & $\it JS_{\it accuracy}$ ($\times 10^{-3}$)& $\it JS_{\it fidelity}$ ($\times 10^{-3}$)\\
		\midrule

		\multirow{4}{*}{CelebA}& Black-box accuracy extraction &1.83 & 9.10\\
		&MH fidelity extraction & 1.42 & 8.17\\
		&Partial black-box  fidelity extraction & 1.53& 8.21\\
		&White-box fidelity extraction & 1.17 & 7.53\\
		
		\cline{1-4}
		\multirow{4}{*}{LSUN-Church}& Black-box accuracy extraction &2.32 & 19.14\\
		&MH fidelity extraction & 2.28 & 19.89\\
		&Partial black-box fidelity extraction & 1.72& 16.95\\
		&White-box fidelity extraction & 1.61 & 18.65\\
		\bottomrule
	\end{tabular}}
\end{table}

\subsection{Defense techniques}
\subsubsection{Returned image visualization for models with defenses techniques.}
\label{ssec:defenses technique}
\begin{figure*}[!t]
\raggedright
  \subfigure[Linear interpolation defense]{
       \includegraphics[width=0.70\linewidth]{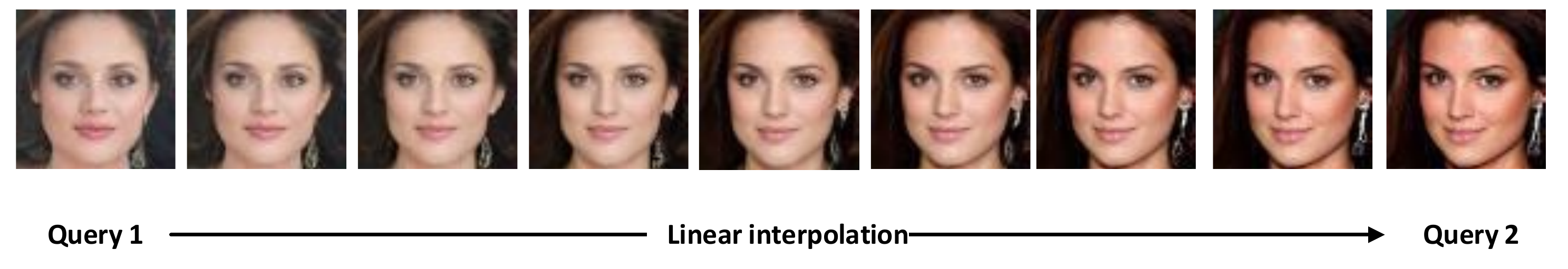}
    \label{fig:linear9}
    }

  \subfigure[Semantic interpolation defense. For one latent code, 12 latent codes containing semantic information are generated through semantic interpolation and corresponding images are shown above.]{
        \includegraphics[width=1\linewidth]{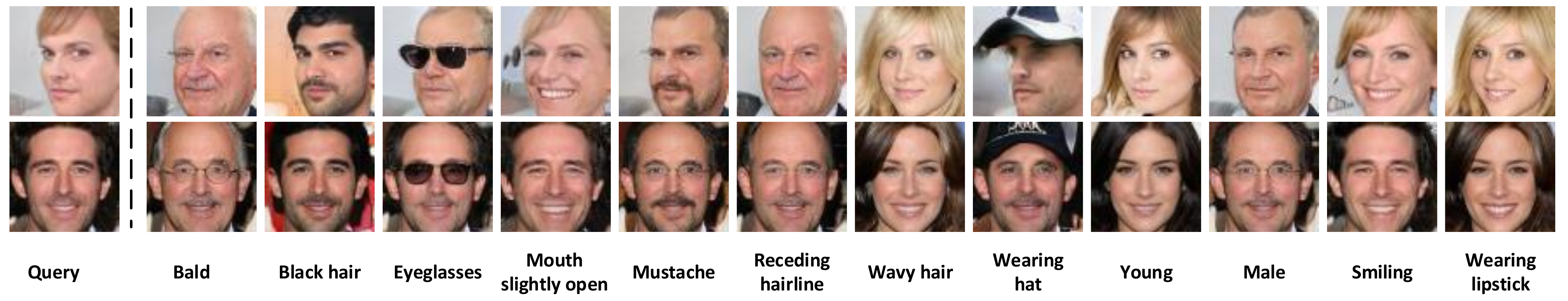}
    \label{fig:sematic}
    }   
    \caption{Returned images after input perturbation-based defense techniques. Queried images and interpolated images both show good quality in visual comparison, and images generated by linear interpolation show more similarity than that by semantic interpolation.}   
    \label{fig:input_defense_images}
\end{figure*}

\begin{figure*}[!ht]
\centering
  \subfigure[Output images. From top to bottom: generated images, Gaussian noise images, Adversarial noise images, Gaussian filter images and JPEG compression images.]{
    \includegraphics[width=1\columnwidth]{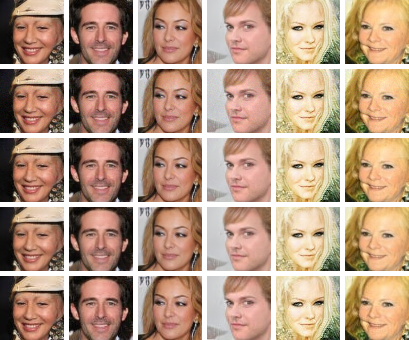}
    \label{fig:output_images}
    }
  \subfigure[Noises. For the top two rows, they are Gaussian noises and adversarial noises, respectively. For the third row, it is the differences between Gaussian filter images and generated images.  For the last row, it is the differences between JPEG compression images and generated images.]{
    \includegraphics[width=1\columnwidth]{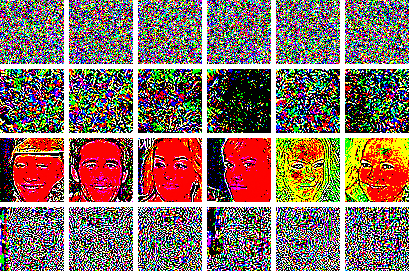}
    \label{fig:output_noise_images}
    }
    \caption{Returned images after output perturbation-based defense techniques.}
    \label{fig:output_defense_images}
\end{figure*}

Figure~\ref{fig:input_defense_images} shows returned images for input perturbation-based defenses.
Figure~\ref{fig:output_defense_images} shows returned images for output perturbation-based defenses.

\subsubsection{Attack performance for the attack model SNGAN}
\begin{figure}[!ht]
\centering
    \includegraphics[width=0.9\columnwidth]{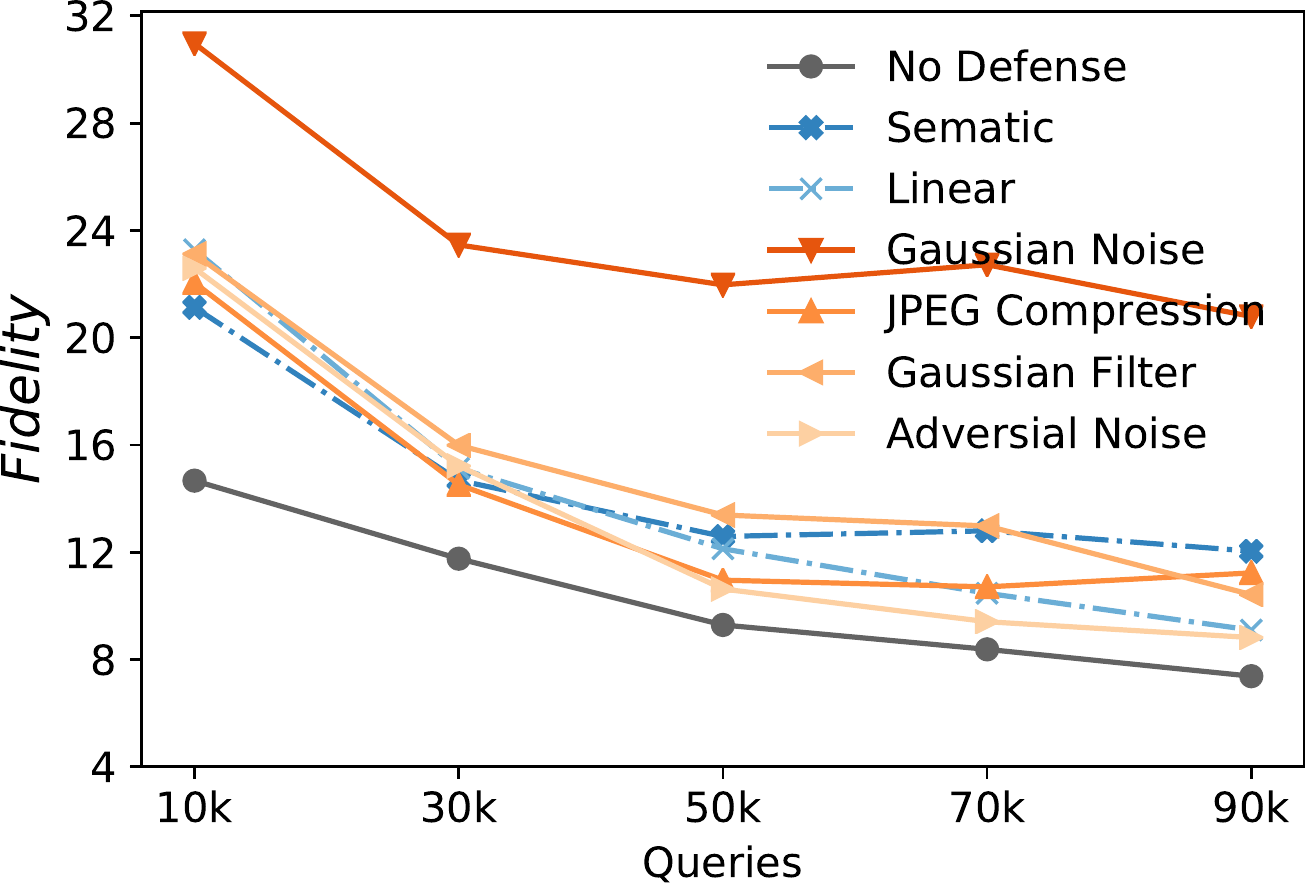}
    \caption{The performance of attack model SNGAN for defenses on black-box accuracy extraction.}
    \label{fig:defense_sngan}
\end{figure}

Figure~\ref{fig:defense_sngan} shows defense performance of the attack model SNGAN for defenses on black-box accuracy extraction. 

\subsubsection{Accuracy on defense techniques}
\label{ssec:Accuracy on defense techniques}
\begin{figure}[!ht]
\centering
  \subfigure[The performance of attack model PGGAN]{
    \includegraphics[width=0.9\columnwidth]{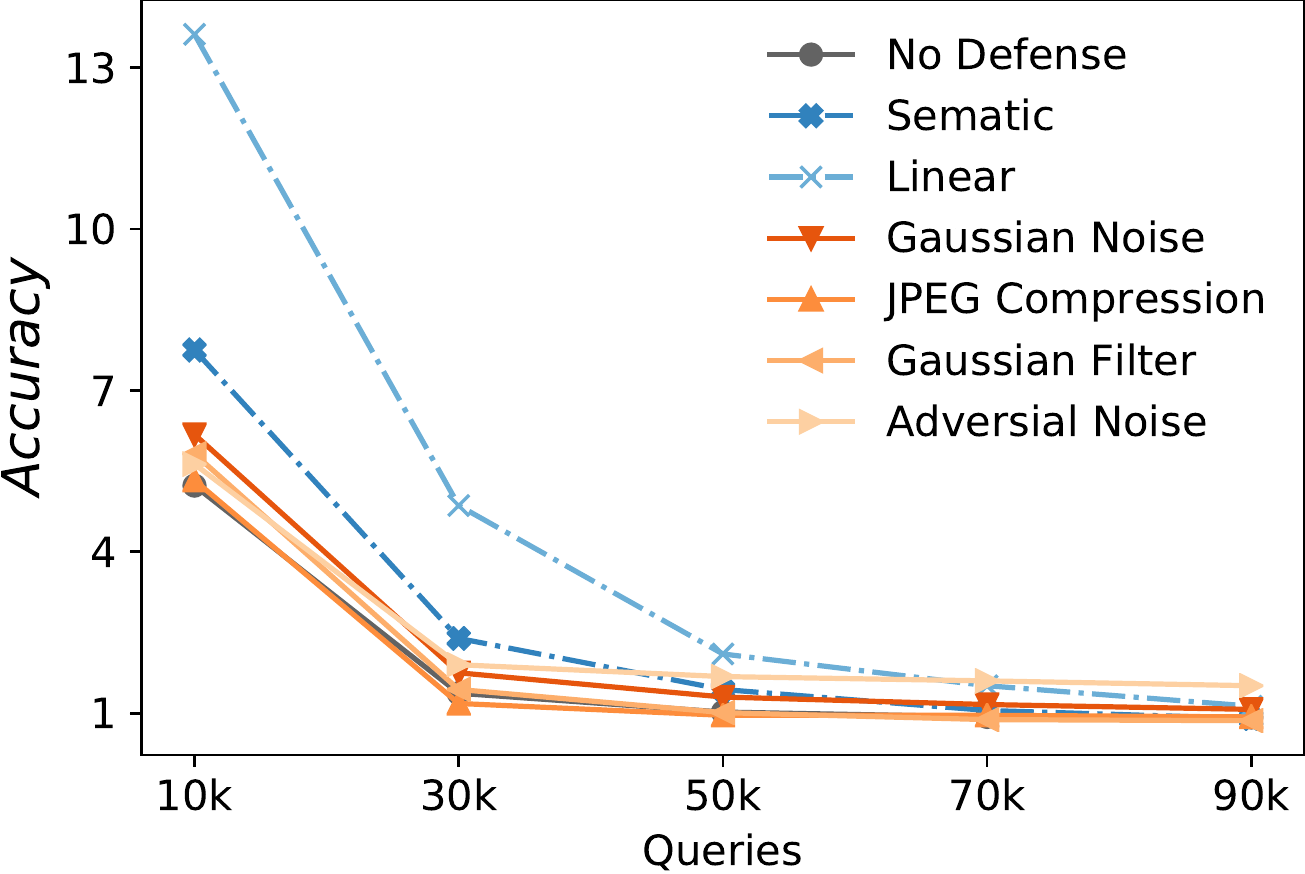}
    \label{fig:defense_pggan_accuracy}
    }
  \subfigure[The performance of attack model SNGAN]{
    \includegraphics[width=0.9\columnwidth]{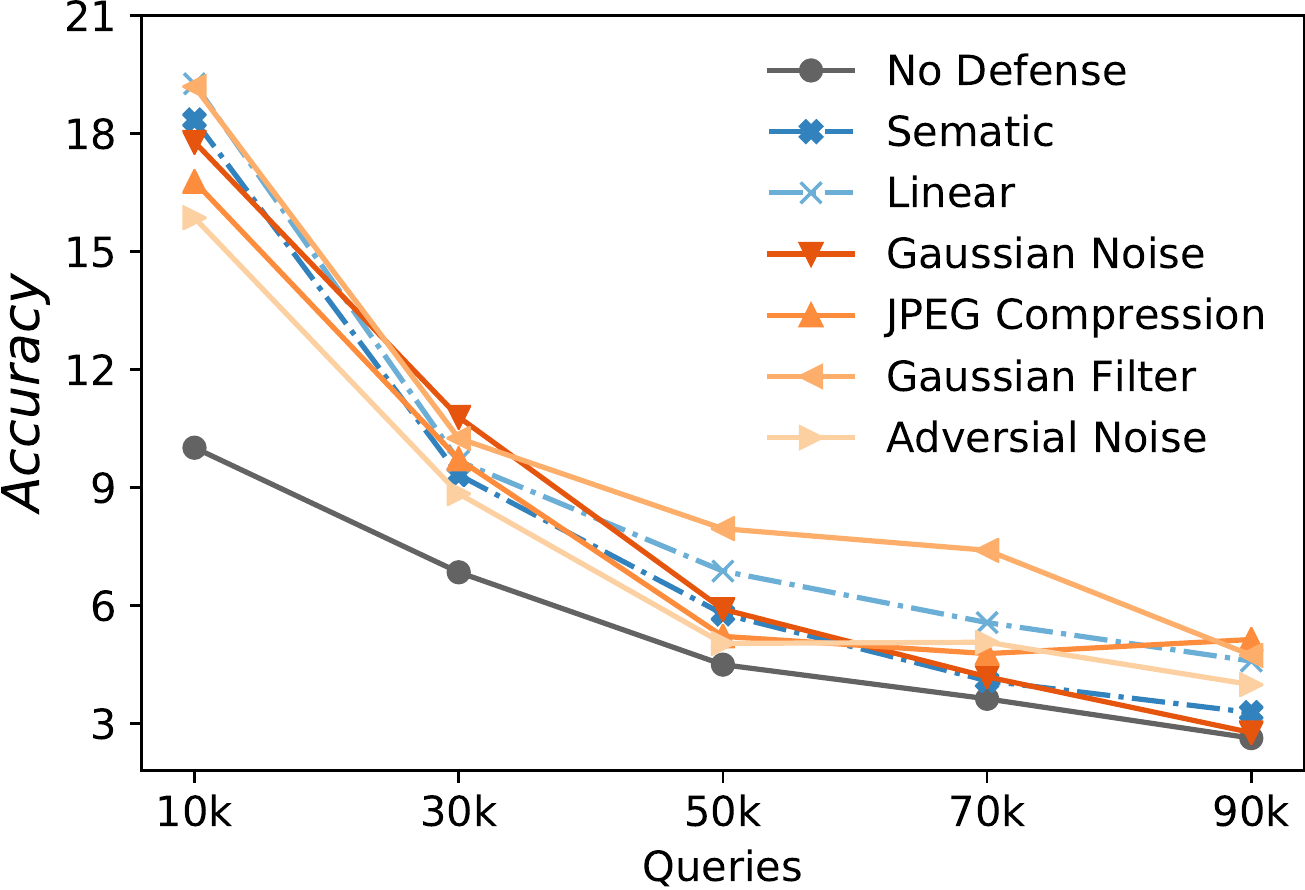}
    \label{fig:defense_sngan_accuracy}
    }
    \caption{{\it Accuracy} of attack models for black-box accuracy extraction. {\it Accuracy} values of attack models can be largely decreased with an increase in the number of queries.}
    \label{fig:Defense_accuracy}
\end{figure}

Figure~\ref{fig:Defense_accuracy} shows $\it accuracy$ of attack models for black-box accuracy extraction. 
We observe that {\it accuracy} values of attack models can be largely decreased with an increase in the number of queries.

\newpage
\subsection{Qualitative results for target models and attack models}
\label{ssec:Qualitative results}
\begin{figure*}[!t]
\centering
    \subfigure{
    \includegraphics[width=0.95\linewidth]{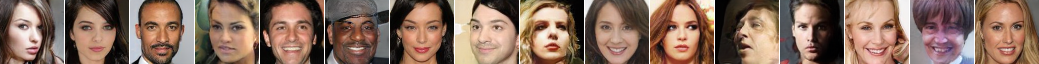}
    \label{fig:pggan_celeba}
    }
  \subfigure{
    \includegraphics[width=0.95\linewidth]{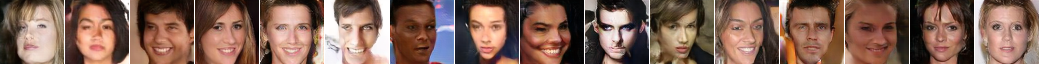}
    \label{fig:sngan_celeba}
    }
  \subfigure{
    \includegraphics[width=0.95\linewidth]{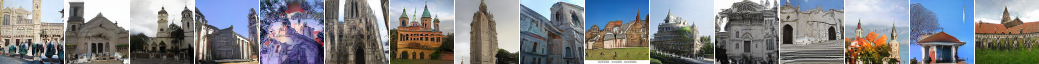}
    \label{fig:pggan_church}
    }
  \subfigure{
    \includegraphics[width=0.95\linewidth]{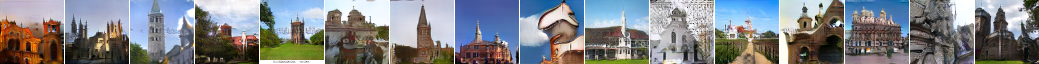}
    \label{fig:sngan_church}
    }
    \caption{Generated images from target GAN models. From top to bottom: PGGAN on CelebA, SNGAN on CelebA, PGGAN on LSUN-Church and SNGAN on LSUN-Church.}
    \label{fig:target_gan}
\end{figure*}

\begin{figure*}[!t]
\centering
    \subfigure{
    \includegraphics[width=0.95\linewidth]{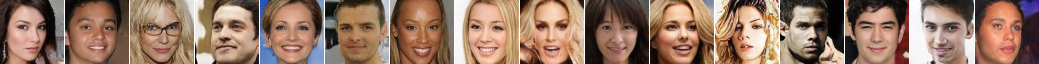}
    \label{fig:pggan_pggan_celeba}
    }
  \subfigure{
    \includegraphics[width=0.95\linewidth]{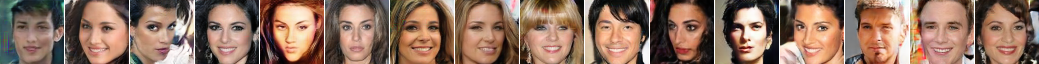}
    \label{fig:sngan_pggan_celeba}
    }
  \subfigure{
    \includegraphics[width=0.95\linewidth]{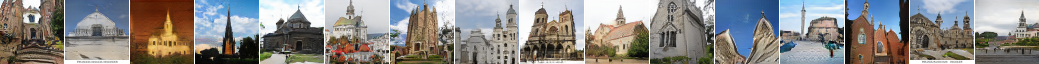}
    \label{fig:pggan_pggan_church}
    }
  \subfigure{
    \includegraphics[width=0.95\linewidth]{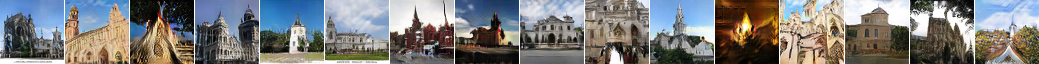}
    \label{fig:sngan_pggan_church}
    }
    
    \subfigure{
    \includegraphics[width=0.95\linewidth]{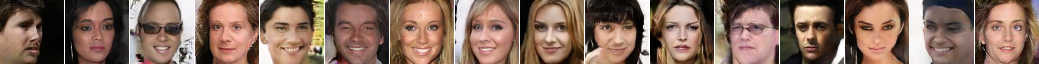}
    \label{fig:pggan_sngan_celeba}
    }
  \subfigure{
    \includegraphics[width=0.95\linewidth]{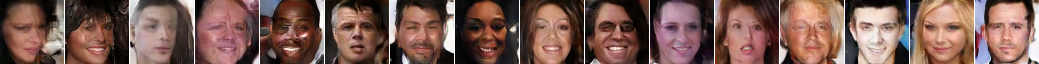}
    \label{fig:sngan_sngan_celeba}
    }
  \subfigure{
    \includegraphics[width=0.95\linewidth]{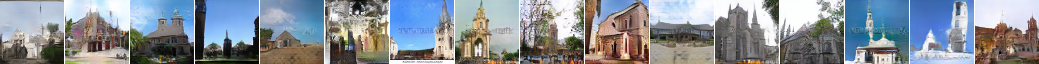}
    \label{fig:pggan_sngan_church}
    }
  \subfigure{
    \includegraphics[width=0.95\linewidth]{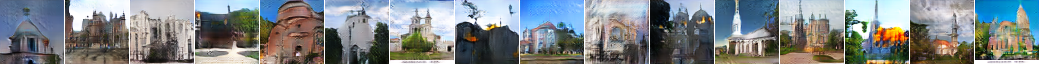}
    \label{fig:sngan_sngan_church}
    }
        
    \caption{The performance of attack models. The first four rows show the performance of attack models against target model PGGAN. From top to the fourth row: PGGAN-PGGAN on CelebA, SNGAN-PGGAN on CelebA, PGGAN-PGGAN on LSUN-Church and SNGAN-PGGAN on LSUN-Church. The last four rows show the performance of attack models against target model SNGAN. From the fifth row to bottom: PGGAN-SNGAN on CelebA, SNGAN-SNGAN on CelebA, PGGAN-SNGAN on LSUN-Church and SNGAN-SNGAN on LSUN-Church.}
    \label{fig:target_pggan}
\end{figure*}

\begin{figure*}[!t]
\centering
  \subfigure{
    \includegraphics[width=0.95\linewidth]{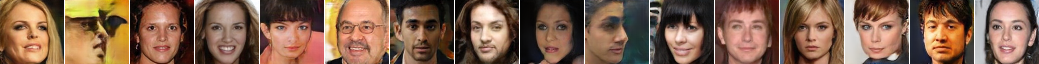}
    \label{fig:pggan_pggan_celeba_linear9}
    }
  \subfigure{
    \includegraphics[width=0.95\linewidth]{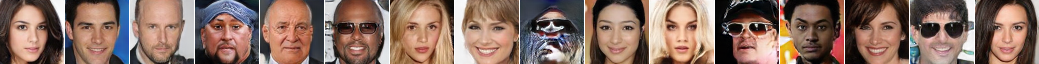}
    \label{fig:pggan_pggan_celeba_semantic}
    }

  \subfigure{
    \includegraphics[width=0.95\linewidth]{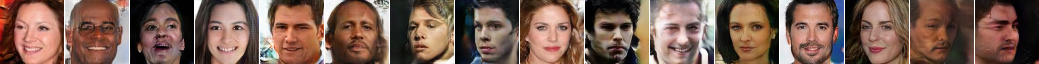}
    \label{fig:sngan_pggan_celeba_linear9}
    }
  \subfigure{
    \includegraphics[width=0.95\linewidth]{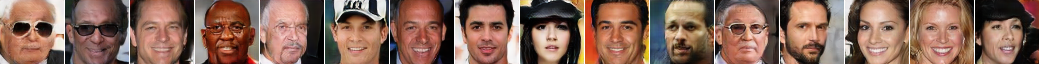}
    \label{fig:sngan_pggan_celeba_semantic}
    }
    
    \caption{The performance of attack models for target model PGGAN with input perturbation-based defense. From top to bottom: PGGAN-PGGAN with linear defense, PGGAN-PGGAN with semantic defense, SNGAN-PGGAN with linear defense, SNGAN-PGGAN with semantic defense.}
    \label{fig:target_pggan_defense_input}
\end{figure*}

\begin{figure*}[!t]
\centering
  \subfigure{
    \includegraphics[width=0.95\linewidth]{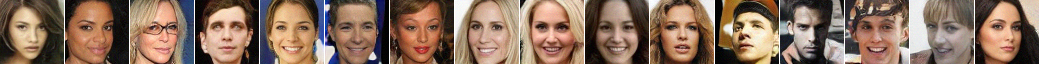}
    \label{fig:defense_PGGAN_gauNoise}
    }
  \subfigure{
    \includegraphics[width=0.95\linewidth]{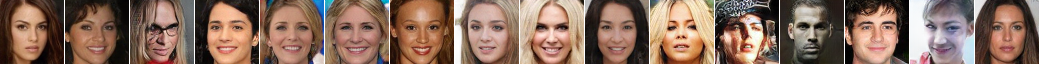}
    \label{fig:defense_PGGAN_advNoise}
    }
  \subfigure{
    \includegraphics[width=0.95\linewidth]{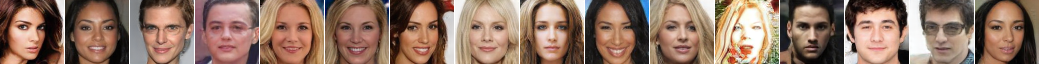}
    \label{fig:defense_PGGAN_gauFilter}
    }
  \subfigure{
    \includegraphics[width=0.95\linewidth]{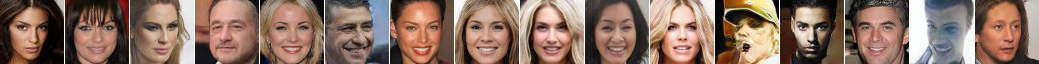}
    \label{fig:defense_PGGAN_compression}
    }

  \subfigure{
    \includegraphics[width=0.95\linewidth]{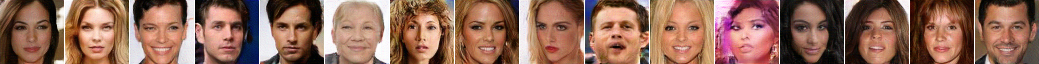}
    \label{fig:defense_SNGAN_gauNoise}
    }
  \subfigure{
    \includegraphics[width=0.95\linewidth]{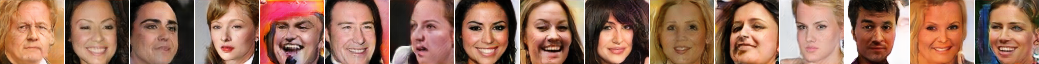}
    \label{fig:defense_SNGAN_advNoise}
    }
  \subfigure{
    \includegraphics[width=0.95\linewidth]{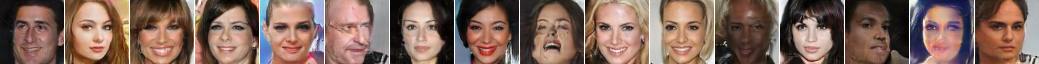}
    \label{fig:defense_SNGAN_gauFilter}
    }
  \subfigure{
    \includegraphics[width=0.95\linewidth]{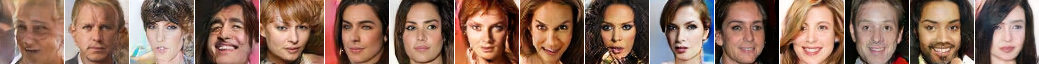}
    \label{fig:defense_SNGAN_compression}
    }
    
    \caption{The performance of attack models for target model PGGAN with output perturbation-based defense. From top to bottom: PGGAN-PGGAN with Gaussian noise defense, PGGAN-PGGAN with adversarial noise defense, PGGAN-PGGAN with Gaussian filtering defense, PGGAN-PGGAN with JPEG compression defense, SNGAN-PGGAN with Gaussian noise defense, SNGAN-PGGAN with adversarial noise defense, SNGAN-PGGAN with Gaussian filtering defense, SNGAN-PGGAN with JPEG compression defense.}
    \label{fig:target_pggan_defense_output}
\end{figure*}

Figure~\ref{fig:target_gan} shows generated images from target GANs. 
Figure~\ref{fig:target_pggan} shows the performance of attack models against target model PGGAN and SNGAN, respectively.  Figure~\ref{fig:target_pggan_defense_input} shows performance of attack models when target model PGGAN trained on CelebA uses input perturbation-based defense. 
Figure~\ref{fig:target_pggan_defense_output} shows performance of attack models when target model PGGAN trained on CelebA uses output perturbation-based defense.

